\documentclass[lettersize,draftcls, onecolumn, 12pt]{IEEEtran}
\usepackage{amsmath,amsfonts}
\usepackage{algorithmic}
\usepackage{algorithm}
\usepackage{array}
\usepackage{subfig}
\usepackage{textcomp}
\usepackage{stfloats}
\usepackage{url}
\usepackage{verbatim}
\usepackage{graphicx}
\usepackage{cite}
\usepackage{epsfig}
\usepackage{epstopdf}
\usepackage{xcolor}
\usepackage{booktabs}
\usepackage{amssymb}
\usepackage{color}
\usepackage{mathrsfs}
\usepackage{array}
\usepackage{amsfonts}
\usepackage{latexsym}
\usepackage{epstopdf}
\usepackage{epsfig}
\usepackage{listings}
\usepackage{pifont}

\usepackage{algorithm}

\usepackage{amsmath}

\usepackage{CJK}
\usepackage{indentfirst}
\usepackage{amsmath}

\usepackage[numbers,sort&compress]{natbib}

\newcommand\myfigpath{fig}

\usepackage[numbers,sort&compress]{natbib}

\def\m #1{\boldsymbol{#1}}

\def\sbra #1{\left(#1\right)}

\def\diag #1{\text{diag}#1}
\def\tr #1{\text{tr}#1}

\def\rank #1{\text{rank}#1}

\newtheorem{lem}{Lemma}
\newtheorem{Remark}{\bf Remark}

\begin{document}
\title{RIS-aided MIMO Beamforming: Piece-Wise Near-field Channel Model}

\author{Weijian Chen, Zai Yang, \IEEEmembership{Senior Member,~IEEE,} Zhiqiang  Wei, \IEEEmembership{Member, IEEE,} Derrick Wing Kwan Ng, \IEEEmembership{Fellow, IEEE,} and Michail Matthaiou, \IEEEmembership{Fellow, IEEE}
        % <-this % stops a space
        
\thanks{Part of the paper has been submitted to the 2024 ICCC \cite{chen2024beamforming}.
	
	W. Chen, Z. Yang, and Z.  Wei are with the School of Mathematics and Statistics, Xi'an Jiaotong University, Xi'an 710049, China (e-mails: chenwj0812@stu.xjtu.edu.cn, yangzai@xjtu.edu.cn, zhiqiang.wei@xjtu.edu.cn). 
	\emph{(Corresponding author: Zhiqiang Wei)}.
	
	D. W. K. Ng is with the School of Electrical Engineering and Telecommunications, the University of New South Wales, Australia (email: w.k.ng@unsw.edu.au).
	
	M. Matthaiou is with the Centre for Wireless Innovation (CWI), Queen's University Belfast, BT3 9DT Belfast, U.K. (e-mail: m.matthaiou@qub.ac.uk).} % <-this % stops a space
	
%\thanks{Author for correspondence: Zhiqiang Wei.}
}

\maketitle
\vspace{-1.6cm}
\begin{abstract}
This paper proposes a joint active and passive beamforming design for reconfigurable intelligent surface (RIS)-aided wireless communication systems, adopting a piece-wise near-field channel model.
While a traditional near-field channel model, applied without any approximations, offers higher modeling accuracy than a far-field model, it renders the system design more sensitive to channel estimation errors (CEEs). 
As a remedy, we propose to adopt a piece-wise near-field channel model that leverages the advantages of the near-field approach while enhancing its robustness against CEEs. Our study analyzes
the impact of different channel models, including the traditional near-field, the proposed piece-wise near-field and far-field channel models, on the interference distribution caused by CEEs and model mismatches. Subsequently, by treating the interference as noise, we formulate a joint active and passive beamforming design problem to maximize the spectral efficiency (SE).
The formulated problem is then recast as a mean squared error (MSE) minimization problem and a suboptimal algorithm is developed to iteratively update the active and passive beamforming strategies.
Simulation results demonstrate that adopting the piece-wise near-field channel model leads to an improved SE compared to both the near-field and far-field models in the presence of CEEs. Furthermore, the proposed piece-wise near-field model 
achieves a good trade-off between modeling accuracy and system's degrees of freedom (DoF).
\end{abstract}


\section{Introduction}
\IEEEPARstart{R}{econfigurable} intelligent surface (RIS)-aided wireless communications have attracted significant interest due to their excellent ability to mitigate the propagation path loss through passive beamforming and to circumvent potential obstacles via establishing alternative propagation paths \cite{matthaiou2021road}. An RIS is a metalic planar array consisting of numerous passive elements that can be independently  reconfigured. By customizing the phase shift of each RIS element based on the channel conditions, the received signal strength at the desired user can be enhanced \cite{cui2014coding}, while undesired interference from adjacent cells and other users can be  efficiently suppressed  \cite{cui2014coding, zhang2020prospective, basar2019wireless, pan2021reconfigurable}.

Numerous studies have focused on optimizing both the active beamforming at the transmitter (Tx) and passive beamforming at the RIS for RIS-aided communication systems \cite{yu2019miso, zhang2020capacity, alwazani2020intelligent, pan2020multicell, hu2021robust, wei2020sum, liu2021deep}. For example, Yu {\it et al.} \cite{yu2019miso} proposed two algorithms for single-user scenarios, namely fixed point iteration and manifold optimization, to maximize the achievable rate for a RIS-aided point-to-point (P2P) multiple-input single-output (MISO) communication system. Similarly, Zhang {\it et al.} \cite{zhang2020capacity} established the fundamental capacity limit through the joint design of the RIS reflection coefficients and the transmit power allocation. On the other hand, in multi-user scenarios, Alwazani {\it et al.} \cite{alwazani2020intelligent} studied the joint optimization of transmit power strategy at the Tx and passive beamforming at the RIS to maximize the minimum user signal-to-interference-plus-noise ratio (SINR), accounting for imperfect channel estimation information (CSI). Additionally, Pan {\it et al.} \cite{pan2020multicell} explored the weighted sum rate of all users in multi-cell scenarios by jointly designing the active and passive beamforming, subject to individual  base station (BS)'s power constraint and the unit modulus constraint for the RIS reflection coefficients according to a weighted minimum mean square error (WMMSE) framework. The affordability and portability of RIS technology have sparked a range of research initiatives. The work of Hu {\it et al.} in \cite{hu2021robust} on multiuser MISO downlink communications utilized an intelligent reflection surface (IRS) capable of reflecting the signal and harvesting energy from signals, enhancing both the communication robustness and security. Also, Wei {\it et al.} explored the implementation of IRS in unmanned aerial vehicle (UAV)-based orthogonal frequency division multiple access (OFDMA) communication systems, utilizing the mobility of UAVs and the beamforming capabilities of the IRS to boost the system's sum-rate \cite{wei2020sum}. Furthermore, deep learning (DL) has been adopted to design beamforming for RIS-assisted multiuser communications, potentially outperforming traditional model-based techniques \cite{liu2021deep}. It is worth noting that these studies assumed a far-field channel model, which is only applicable when the RIS is deployed far from both the Tx and users.

In practice, the electromagnetic (EM) radiation field is commonly divided into two regions: the far-field and the near-field. The demarcation between these regions is determined by the Rayleigh distance which is proportional to the square of the array aperture and is inversely proportional to the signal carrier wavelength \cite{cui2022near}. Specially, when a receiver (Rx) is located in the far-field region, the EM field propagation is approximately modeled by planar waves with a certain approximate error. However, in the near-field region, near-field propagation becomes dominant and the EM field propagation is accurately modeled by spherical waves. In fact, with emerging high-frequency communications, utilizing large-aperture antenna arrays for potential power gains can also increase the Rayleigh distance up to a hundred meters, such that the far-field assumptions are no longer valid.

In recent years, a novel area of research has emerged with the goal of improving the efficiency and performance of near-field wireless communication systems. For instance, in \cite{cui2022near}, Cui {\it et al.} introduced the fundamental concept of RIS-assisted near-field communications and highlighted several areas for future research. Besides, Wei {\it et al.} \cite{wei2022codebook} focused on developing a near-field codebook for extremely large-scale RIS (XL-RIS) by taking into account the near-field cascaded array steering vector. Specifically, they crafted a hierarchical near-field codebook and introduced a corresponding hierarchical near-field beam training scheme to minimize the beam training overhead.
 Furthermore, Wang {\it et al.} \cite{wang2024base} proposed two efficient schemes for optimizing the BS beamforming to improve the RIS beam training performance, leveraging a near-field channel model. Moreover, in \cite{dovelos2021intelligent}, Dovelos {\it et al.} investigated RIS-aided MIMO systems in terms of power gain and energy efficiency (EE), considering a spherical wave channel model. Specifically, they analyzed the power gains under beamfocusing and beamsteering, concluding that beamfocusing in the radiating near-field is more useful than beamforming adopted in the far-field counterpart scenarios.

In fact, the necessity of adopting a near-field channel model in a RIS-aided wireless communication system is twofold. Firstly, the RIS is usually coated with a large aperture and needs to be deployed close to the Tx or the Rx to enhance its gain \cite{tao2020performance}, \cite{li2021intelligent}. In other words, the Tx or Rx is more likely to be within the near-field region of the RIS. Secondly, when the RIS is located close to the Tx or the Rx, either the Tx-RIS or the RIS-Rx channel is likely dominated by line-of-sight (LoS) propagation paths \cite{liu2023near}, \cite{wei2021channel}. Note that assuming a LoS far-field channel in either the Tx-RIS or the RIS-Rx link makes the cascaded channel matrix between the Tx and Rx only exhibits rank-one and, thus, limits the system's design degrees of freedom (DoF). It is important to note that a practical near-field channel model, as discussed in \cite{zhou2015spherical}, can effectively uncover the implicit higher ranks of the cascaded channel matrix, potentially leading to an improvement in both the system's DoF and spectral efficiency (SE). As a result, considering a near-field channel model for RIS-aided wireless communication systems not only improves the modeling accuracy but also reveals more DoF.

Despite of the advantages mentioned above, embracing near-field communications entails significant challenges. One key challenge is that characterizing a near-field channel requires knowledge of all the distances between each pair of transceiver antennas, whereas the far-field channel only depends on the distance between transceivers, the angle of departure (AoD) at the Tx, and the angle of arrival (AoA) at the Rx. Consequently, estimating the near-field channel is generally more challenging due to the larger number of parameters involved. Another challenge is the sensitivity of near-field communications to channel estimation errors (CEEs). For a far-field channel model, aligning the beam direction of the Tx towards the Rx is usually sufficient to achieve the beamforming gain, as long as the Rx is within the beam's cone. However, for near-field communications, precise beamfocusing is required to accurately focus the transmitted signal towards the Rx's location. In practical situations with the presence of CEEs, the focus of near-field beamforming may deviate from the intended focal point, which degrades the beamforming gain and system performance \cite{wei2021channel}. These two challenges motivate us to consider a new channel model that not only leverages the advantages of the near-field channel model but also enhances its robustness against CEEs.

%For a better understanding of our contributions, we have presented a comparison table with existing works. 
\begin{table}[t] 
	\centering 
	\label{Compare} 
	\caption{Comparison of Existing Works} 
	%\vspace{5pt} 
	\begin{tabular}{ccccc} 
		\toprule
		References &Joint Beamforming  & MIMO & Channel Model & CSI Assumption \\ 
		\midrule
		\cite{yu2019miso,alwazani2020intelligent,hu2021robust,wei2020sum,liu2021deep} & \checkmark &\ding{55} & Far-field &Perfect/Imperfect \\
		\cite{zhang2020capacity,pan2020multicell} &\checkmark &\checkmark &Far-field &Perfect \\
		\cite{cui2022near}, \cite{wei2022codebook, wang2024base, dovelos2021intelligent} &\checkmark &\checkmark & Near-field &Perfect \\
		This paper &\checkmark &\checkmark &Piece-wise near-field  & Imperfect \\
		\bottomrule
	\end{tabular}
\end{table}

As compared in Table I, various existing works have overlooked the near-field effect and some recent works on near-field communications assumed the ideal case of perfect CSI for beamforming design. To the best of the authors' knowledge, our work is the first to compare the performance and robustness of different channel models for RIS-aided communications. 

In comparison to the conference version \cite{chen2024beamforming}, this paper provides a detailed description of the piece-wise near-field model and extensively discusses the rank of the cascaded channel and its advantages. Additionally, it presents a comprehensive algorithmic framework, parameter selection, and convergence description for the the alternating direction penalty method (ADPM) algorithm, addressing the RIS phase constant modulus constraint. Through simulations, this paper also illustrates the impact of the number of antennas at the Tx and the number of RIS reflecting elements on SE, shedding light on their distinct roles in RIS-aided communication systems. The key contributions of this paper are outlined as follows:
\begin{itemize}
\item {For the first time, we propose a joint active and passive beamforming design for RIS-aided MIMO wireless communication systems, adopting a piece-wise near-field channel model. This model approximates the near-field channel via dividing a large-aperture RIS into multiple small-aperture sub-surfaces and assuming heterogeneous far-field propagation between each surface to the Tx or Rx. One can imagine that the piece-wise near-field channel model retains the advantages of accurate channel modeling and increased DoF gain of the near-field channel model.} %while reducing the number of parameters involved and might improve the system design robustness. }

\item{For the three different channel models, i.e., the near-field model, the far-field model and the piece-wise near-field model, and by assuming an identical normalized CEE but different model mismatches, we analyze the covariance matrices of the corresponding interference plus noise signals.} 

\item The joint active and passive beamforming design is formulated as an optimization problem to maximize the achievable SE by treating the interference caused by CEEs and model mismatches as noise. The achievable SE maximization problem is then transformed equivalently to a problem minimizing the mean square error (MSE), assuming a Gaussian CEE distribution. 

\item A block coordinate descent (BCD) approach is adopted to alternately optimize the active and passive beamforming strategies. In particular, the ADPM is adopted to address the constant modulus constraint on the RIS reflection coefficient. We propose an iterative suboptimal algorithm with closed-form updating rules in each step to design the active and passive beamforming strategies. Simulation results demonstrate the system design DoF gain and the enhanced robustness against CEE of adopting the piece-wise near-filed channel model compared to the conventional far-field and near-field models.
\end{itemize}

%6. Organization of this work.
The rest of this paper is organized as follows:  In  Section II,  we introduce the RIS-aided near-field communication system model adopting the piece-wise near-field channel model. Section III provides the analysis of the interference distribution with different channel models. In Section IV, we formulate the joint active and passive beamfoming design problem. Then, we present the solution in Section V. In Section VI, we present numerical results with discussions. Finally, we conclude with Section VII.

\emph{Notations:} Lower-case letters are used to represent scalars, while vectors and matrices are denoted by lower-case and upper-case boldface letters, respectively. The set of complex numbers is denoted by $\mathbb{C}$; $\Re$ extracts the real part of a complex number. For vector $\boldsymbol{x}$, $\boldsymbol{x}_j$ denotes the $j$-th element of $\m{x}$ and $\diag\sbra{\m{x}}$ denotes a diagonal matrix with its diagonal entries given by ${\m{x}}$. For matrix $ \boldsymbol{A}$,  $\mathbb{E}\{\m A\}$, $ \boldsymbol{A}^{T} $, $ \boldsymbol{A}^{H} $, $ \|\boldsymbol{A}\|_F $, $ \boldsymbol{A}^{-1} $, $\rank\sbra{\m{A}}$, and $\tr\sbra{\m{A}}$ denote the expectation, matrix transpose, conjugate transpose, Frobenius norm, inverse, rank, and trace of $ \boldsymbol{A}$, respectively; $\mathcal{CN}(\boldsymbol{\mu}, \boldsymbol{\Sigma})$ denotes a circularly symmetric complex Gaussian random vector distribution with mean $\boldsymbol{\mu}$ and covariance matrix $\boldsymbol{\Sigma}$. The matrix $\m A$ is said to have a matrix-variate complex Gaussian distribution, which can be written as $\mathcal{CN}_{n,m}(\boldsymbol{\Pi}_{n\times m}, \boldsymbol{\Sigma}_{\rm r}\otimes \boldsymbol{\Sigma}^T_{\rm c})$, where the $n \times n$ matrix $\boldsymbol{\Sigma}_{\rm r}$ and the $m \times m$ matrix $\boldsymbol{\Sigma}_{\rm c}$ are the
row and column covariance matrices of $\boldsymbol{A}$, respectively \cite{gupta2018matrix}.  Besides, $ \rm vec (\boldsymbol{A}) \sim \mathcal{CN}(\rm vec ({\m \Pi}), \boldsymbol{\Sigma}_r \otimes \boldsymbol{\Sigma}^T_c) $ where the operation $\rm vec(\boldsymbol{A})$ stacks the columns of the matrix $\boldsymbol{A}$ into a single vector. The symbols $\otimes$ and $\circ$ represent the Kronecker and Hadamard products, respectively. The symbol $\angle$ is used to denote the phase of a complex number. 

\begin{figure}[t]
	\centering
	\subfloat[Near-field channel model for the Tx-RIS channel.]{
		\hspace{-1.cm} \includegraphics[scale=0.26]{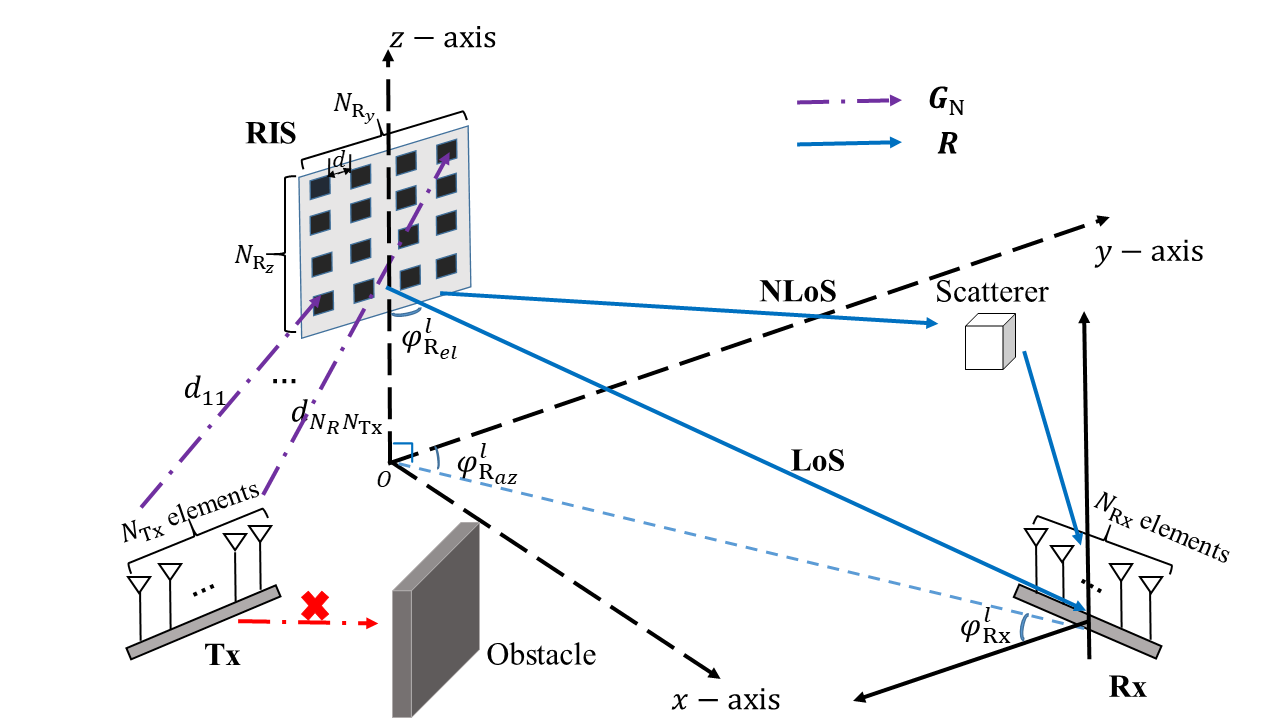}}
	\subfloat[Piece-wise near-field channel model for the Tx-RIS channel.]{
		\includegraphics[scale =0.26]{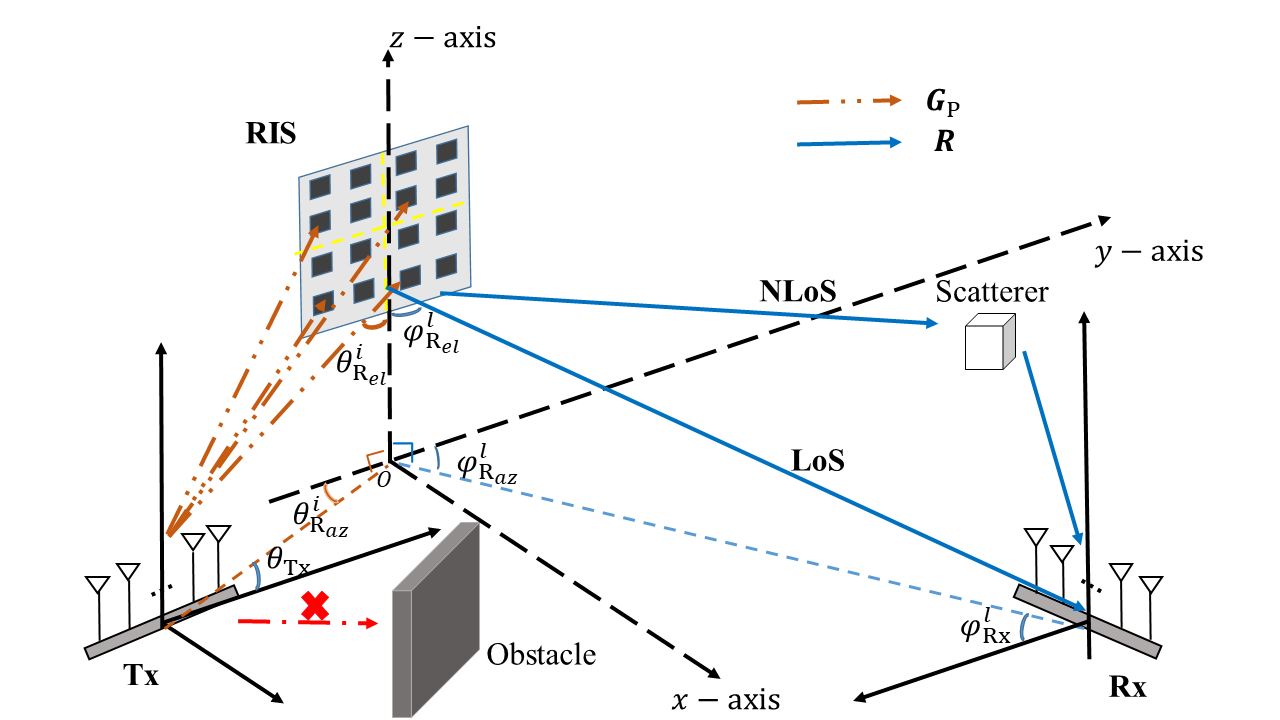}}
	\caption{The RIS-aided P2P wireless communication system.}
	\label{fig1}
\end{figure}

\section{System Model}
%1) System model Figure to show the system setup
\subsection{System Model}
We consider a  RIS-aided point-to-point (P2P) multiple-input multiple-output (MIMO) wireless communication system, as illustrated in Fig. 1, where the RIS is deployed close to a Tx to assist the data transmission from the Tx to a Rx.  The Tx equipped with ${N_{\rm Tx}}$ antennas, transmits ${N_{\rm s}} \ {(\leq N_{\rm Tx})}$ independent data streams to the Rx equipped with $N_{\rm Rx}$ antennas with the aid of a RIS comprising $N_{\rm R}=N_{{\rm R}_y}\times N_{{\rm R}_z}$ passive elements, where $N_{{\rm R}_y}$ and $N_{{\rm R}_z}$ are the number of elements in the horizontal and vertical directions of RIS, respectively. Without loss of generality, we assume that both the transmit and receive antenna arrays are uniform linear arrays (ULAs) with an identical antenna spacing for both arrays at the Tx and Rx as well as the RIS, denoted by $d$. We assume that there is no direct link from the Tx to the Rx, which is likely to occur due to blockages as commonly assumed in the literature \cite{sid2022reconfigurable}. 
Furthermore, the narrowband channels from the Tx to the RIS and from the RIS to the Rx are denoted by $\boldsymbol{G} \in {\mathbb{C}^{N_{\rm R}\times {N_{\rm Tx}}}}$ and
	$\boldsymbol{R} \in {\mathbb{C}^{N_{\rm Rx}\times {N_{\rm R}}}}$, respectively. The phase shift matrix of the RIS is denoted by  $\boldsymbol{\Phi} = \diag(\phi_1, \phi_2,..., \phi_{N_{\rm R}})\in {\mathbb{C}^{N_{\rm R}\times {N_{\rm R}}}}$,
to perform passive beamforming, where $\phi_{n_{\rm R}} =e^{j{\zeta}_{n_{\rm R}}} $ and $\zeta_{n_{\rm R}}\in [0, 2\pi]$ is the phase shift introduced by the ${n_{\rm R}}$-th RIS element, $\forall {n_{\rm R}} \in \{1,\ldots, N_{\rm R}\} $. We denote the symbols transmitted from the Tx to the Rx as $\boldsymbol{s} \in {\mathbb{C}^{N_{\rm s}\times 1}} \sim \mathcal{CN}(\m{0}, \m{\rm I}_{N_{\rm s}}) $. The received signal $\boldsymbol{y} \in {\mathbb{C}^{N_{\rm Rx}\times 1}}$ at the Rx is given by
	\begin{equation} \label{signalmodel}
		\boldsymbol{y}= \boldsymbol{R} \boldsymbol{\Phi} \boldsymbol{G} \boldsymbol{W} \boldsymbol{s} + \boldsymbol{n}=\boldsymbol{H}\boldsymbol{W} \boldsymbol{s} + \boldsymbol{n},
	\end{equation}
	where  $\boldsymbol{W} \in \mathbb{C}^{N_{\rm Tx}\times {N_{\rm s}}}$ is the precoding matrix (i.e., active beamforming matrix) at the Tx and $\boldsymbol{n} \in \mathbb{C}^{N_{\rm Rx}\times 1}\sim\mathcal{CN}(\m{0}, \sigma^2 \boldsymbol{\rm I}_{N_{\rm Rx}})$ is the additive white Gaussian noise (AWGN) at the Rx with a noise power of $\sigma^2$. For the sake of presentation, we define the cascaded channel between the Tx and Rx as $\boldsymbol{H}=\boldsymbol{R} \boldsymbol{\Phi} \boldsymbol{G}$.

\subsection{Channel Models}
To achieve a large passive beamforming gain, the RIS is deployed close to the Tx as shown in Fig. 1(a) and the number of RIS elements is usually large \cite{cui2022near}, \cite{wei2022codebook}. Therefore, we consider a near-field channel model for the Tx-RIS channel and a far-field multi-path channel model for the RIS-Rx channel. To avoid confusion, we adopt different channel models, i.e., $\m G_{\rm N}, \m G_{\rm P}$, and $\m G_{\rm F}$, to represent the link between Tx-RIS. In particular, assuming a LoS-dominated propagation between the Tx and RIS, the Tx-RIS channel is modeled as  \cite{bjornson2020power}
	\begin{equation}\label{near-field}
		\begin{array}{c}
			\boldsymbol G_{\rm N}
		\end{array}
		=
		\begin{bmatrix}
			\alpha_{11} e^{-j\frac{2\pi}{\lambda}{d_{11}}} & \cdots & \alpha_{1 {N_{\rm Tx}}}e^{-j\frac{2\pi}{\lambda}{d_{1{N_{\rm Tx}}}}}\\
			\alpha_{21} e^{-j\frac{2\pi}{\lambda}{d_{21}}}  & \cdots & \alpha_{2 {N_{\rm Tx}}}e^{-j\frac{2\pi}{\lambda}{d_{2{N_{\rm Tx}}}}}\\
			\vdots & \ddots & \vdots \\
		\alpha_{{N_{\rm R}1}} e^{-j\frac{2\pi}{\lambda}{d_{{N_{\rm R}1}}}} & \cdots & \alpha_{N_{\rm R} N_{\rm Tx}}e^{-j\frac{2\pi}{\lambda}{d_{{N_{\rm R}}{N_{\rm Tx}}}}}
		\end{bmatrix},
	\end{equation}
	where $\alpha_{n_{\rm R} n_{\rm Tx}}=\frac{\lambda^2}{(4\pi d_{n_{\rm R} n_{\rm Tx}})^2}$ is the associated path coefficient, $\lambda$ is wavelength of the signal carrier frequency, and $d_{{n_{\rm R}}{n_{\rm Tx}}}$ is the distance between the ${n_{\rm Tx}}$-th antenna at the Tx and the ${n_{\rm R}}$-th element at the RIS. In contrast, the Tx-RIS channel has been approximated by a far-field channel model in the literature \cite{pan2022overview}--\cite{wang2020channel}, i.e., 
	\begin{equation}\label{far-field}
		\boldsymbol G_{\rm F} = \gamma \m{a}_{N_{\rm R}}(\psi_{{\rm R}_{\rm az}}, \psi_{{\rm R}_{\rm el}}) \m{a}^H_{N_{\rm Tx}}(\theta_{\rm Tx}),
	\end{equation}
	where $\gamma=\frac{\lambda^2}{(4\pi d_{\rm TR})^2} e^{-j\frac{2 \pi}{\lambda}d_{\rm TR}}$ is the path coefficient and $d_{\rm TR}$ is the distance between the centers of Tx and RIS. Moreover, $\m{a}_{N_{\rm Tx}}(\theta_{\rm Tx})\in \mathbb{C}^{N_{\rm Tx}\times 1}$ and $\m{a}_{N_{\rm R}}(\psi_{{\rm R}_{\rm az}}, \psi_{{\rm R}_{\rm el}})\in \mathbb{C}^{N_{{\rm R}}\times 1}$ denote the array response vectors at the Tx and RIS, respectively, which are given by
	\begin{equation} \label{aNTx}
		\m{a}_{N_{\rm Tx}}(\theta_{\rm Tx}) = \left[1,\cdots,e^{-j{\frac{2\pi}{\lambda} (N_{\rm Tx} -1)d \sin{\theta}_{\rm Tx}}}\right]^T,  %\text{ and} 
	\end{equation}
	and
	\begin{align}
		\m{a}_{N_{\rm R}}(\psi_{{\rm R}_{\rm az}}, \psi_{{\rm R}_{\rm el}}) = \left[1,\cdots,e^{-j{\frac{2\pi}{\lambda} (N_{{\rm R}_y} -1)d \sin{\psi}_{{\rm R}_{\rm az}}\cos{\psi}_{{\rm R}_{\rm el}}}}\right]^T 
		\otimes \left[1,\cdots,e^{-j{\frac{2\pi}{\lambda} (N_{{\rm R}_z} -1)d \sin{\psi}_{{\rm R}_{\rm az}}\sin{\psi}_{{\rm R}_{\rm el}}}}\right]^T,
	\end{align}
	respectively, where $\theta_{\rm Tx}$ is the azimuth AoD from the Tx to the center of the RIS, while $\psi_{{\rm R}_{\rm az}}, \psi_{{\rm R}_{\rm el}}$ are the azimuth and elevation AoAs at the center of the RIS. 
	
	On the other hand, based on the Saleh-Valenzuela channel model \cite{el2014spatially}, the RIS-Rx channel matrix is given by
	\begin{equation}\label{far}
		\boldsymbol{R} = \sum_{l=1}^{L_{\rm Rx}} \beta_l {\boldsymbol{a}_{N_{\rm Rx}}}({\varphi}_{{\rm Rx}}^l) \boldsymbol{b}_{N_{\rm R}}^H({\varphi}_{{\rm R}_{\rm az}}^l,{\varphi}_{{\rm R}_{\rm el}}^l),
	\end{equation}
	where $\beta_l=\frac{\lambda^2}{(4\pi d_{\rm RR})^2}e^{-j\frac{2 \pi}{\lambda}d_{\rm RR}}$ is the path coefficient of the $l$-th path between the RIS and Rx, $d_{\rm RR}$ is the distance between the centers of RIS and Rx, and $L_{\rm Rx}$ is the total number of paths. The vectors ${\boldsymbol{a}_{N_{\rm Rx}}}({\varphi}_{{\rm Rx}}^l)\in \mathbb{C}^{N_{{\rm Rx}}\times 1}$ and ${\boldsymbol{b}_{N_{\rm R}}}({\varphi}_{{\rm R}_{\rm az}}^l,{\varphi}_{{\rm R}_{\rm el}}^l)\in \mathbb{C}^{N_{\rm R}\times 1}$ denote the array response vectors at the Rx and RIS, respectively, and they are given by 
	\begin{equation} 
		{\boldsymbol{a}_{N_{\rm Rx}}}({\varphi}_{\rm Rx}^l)=\left[1,  \cdots, e^{-j{\frac{2\pi}{\lambda} (N_{\rm Rx} -1)d \sin{\varphi}_{\rm Rx}^l}}\right]^T %{\text { and}}
	\end{equation}
	and
	\begin{align}
		{\boldsymbol{b}_{N_{\rm R}}}({\varphi}_{{\rm R}_{\rm az}}^l,{\varphi}_{{\rm R}_{\rm el}}^l)=\left[1, \cdots, e^{-j{\frac{2\pi}{\lambda} (N_{{\rm R}_y} -1)d \sin{\varphi}_{{\rm R}_{\rm az}}^l \cos{\varphi}_{{\rm R}_{\rm el}}^l}}\right]^T  \otimes \left[1, \cdots, e^{-j{\frac{2\pi}{\lambda} (N_{{\rm R}_z} -1)d \sin{\varphi}_{{\rm R}_{\rm az}}^l \sin{\varphi}_{{\rm R}_{\rm el}}^l}}\right]^T,
	\end{align}
	respectively, where ${\varphi}_{\rm Rx}^l$ is the azimuth AoA of the $l$-th path at the Rx, and ${\varphi}_{{\rm R}_{\rm az}}^l$ and ${\varphi}_{{\rm R}_{\rm el}}^l$ are the azimuth and elevation AoDs of the $l$-th path from the RIS to the Rx, respectively.
	
	Comparing (\ref{near-field}) and (\ref{far-field}), we can observe that the near-field channel model in (\ref{near-field}) involves more parameters than that in (\ref{far-field}). Indeed, when the Tx is located within the near-field region of the RIS, (\ref{near-field}) is a more accurate model to describe the signal propagation between the Tx and RIS, compared to (\ref{far-field}). It has been demonstrated in \cite{zhou2015spherical}, \cite{liu2023near}, and \cite{zhao20246g} that $\rank({\m{G}_{\rm N}})>\rank({\m{G}_{\rm F}})=1$ usually holds when the near-field condition is satisfied, i.e., the distance between the Tx and RIS is shorter than the Rayleigh distance. However, for the near-field channel model in (\ref{near-field}), beamfocusing is required for passive beamforming design at the RIS, which is typically more sensitive to CEEs than conventional beamsteering for the far-field channel model. Therefore, we advocate the utilization of a piece-wise near-field channel model to approximate the near-field channel in (\ref{near-field}). We propose to equally divide the $N_{{\rm R}_y}$ RIS elements in each row of the RIS into $K$ subarrays, and divide the $N_{{\rm R}_z}$ RIS elements in each column of the RIS into $K$ subarrays. Without lost of generality, we assume that both $\frac{N_{{\rm R}_y}}{K}$ and $\frac{N_{{\rm R}_z}}{K}$ are integers. As a result, the original RIS is divided into $K^2$ subsurfaces and the Rayleigh distance is reduced by a factor of $K^2$. Consequently, we can safely assume that the channel between each subsurface and the Tx follows a far-field channel model.
	Thus, the proposed piece-wise near-field channel matrix, $ \boldsymbol{G}_{\rm P}$, is given by
	\begin{equation}\label{Mblock}
		\m{G}_{\rm P}=
		\sbra{\begin{bmatrix}
				\m{g}^{\rm h}_1\\
				\vdots \\
				\m{g}^{\rm h}_{K}\\
			\end{bmatrix}
			\otimes
			\begin{bmatrix}
				\boldsymbol g^{\rm v}_{1}\\
				\vdots \\
				\m{g}^{\rm v}_{K}\\
		\end{bmatrix}}
		{\boldsymbol{a}_{N_{\rm Tx}}^H}({\theta}_{\rm Tx}),
	\end{equation}
	where ${\boldsymbol{a}_{N_{\rm Tx}}}({\theta}_{\rm Tx})$ was defined in (\ref{aNTx}), while $\m{g}^{\rm h}_i \in \mathbb{C}^{{\frac{N_{{\rm R}_y}}{K}} \times 1}$ and $\m{g}^{\rm v}_i \in \mathbb{C}^{{\frac{N_{{\rm R}_z}}{K}} \times 1}$ are defined by
	\begin{equation}\label{Mblockx}
		\boldsymbol{g}^{\rm h}_{i} = \frac{\lambda}{4 \pi r_{i}} e^{-j{\frac{\pi}{\lambda} r_{i}}} \boldsymbol{b}_{\frac{N_{{\rm R}_y}}{K}}(\theta^{i}_{{\rm R}_{az}}, \theta^{i}_{{\rm R}_{el}}), {i}=1,\ldots,K, %\text{ and}
	\end{equation}
	and
	\begin{equation}\label{Mblocky}
		\boldsymbol{g}^{\rm v}_{i} = \frac{\lambda}{4 \pi r_{i}} e^{-j{\frac{\pi}{\lambda} r_{i}}} \boldsymbol{b}_{\frac{N_{{\rm R}_z}}{K}}(\theta^{i}_{{\rm R}_{az}}, \theta^{i}_{{\rm R}_{el}}), {i}= 1,\ldots, K,
	\end{equation}
	respectively. In (\ref{Mblockx}) and (\ref{Mblocky}), ${r_i}$, ${\theta}_{{\rm R}_{\rm az}}^i$, and ${\theta}_{{\rm R}_{\rm el}}^i$ are the distance, azimuth, and elevation AoAs from the Tx to the center of the $i$-th subsurface, respectively. The vectors $\boldsymbol{b}_{\frac{N_{{\rm R}_y}}{K}}({\theta}_{{\rm R}_{\rm az}}^i,{\theta}_{{\rm R}_{\rm el}}^i) $ and ${\boldsymbol{b}_{\frac{N_{{\rm R}_z}}{K}}}({\theta}_{{\rm R}_{\rm az}}^i,{\theta}_{{\rm R}_{\rm el}}^i)$ are the horizontal and vertical array response vectors of the $i$-th subsurface which are given by
	\begin{equation}
		{\boldsymbol{b}_{\frac{N_{{\rm R}_y}}{K}}}({\theta}_{{\rm R}_{\rm az}}^i,{\theta}_{{\rm R}_{\rm el}}^i)=\left[1, \cdots, e^{-j{\frac{2\pi}{\lambda} ({\frac{N_{{\rm R}_y}}{K}} -1)d \sin{\theta}_{{\rm R}_{\rm az}}^i \cos{\theta}_{{\rm R}_{\rm el}}^i}}\right]^T, %\text{ and}
	\end{equation}
	 and
	\begin{equation}
		{\boldsymbol{b}_{\frac{N_{{\rm R}_z}}{K}}}({\theta}_{{\rm R}_{\rm az}}^i,{\theta}_{{\rm R}_{\rm el}}^i)=\left[1, \cdots, e^{-j{\frac{2\pi}{\lambda} ({\frac{N_{{\rm R}_z}}{K}} -1)d \sin{\theta}_{{\rm R}_{\rm az}}^i \sin{\theta}_{{\rm R}_{\rm el}}^i}}\right]^T, 
	\end{equation}
	respectively. For illustration, let us take  $K=2$ as an example, which is shown in Fig. 1(b). In this case, the piece-wise near-field channel matrix $ \boldsymbol{G}_{\rm P}$ is given by 
	\begin{equation}\label{fourblock}
		\m{G}_{\rm P}=
		\sbra{\begin{bmatrix}
				\m{g}^{\rm h}_1\\
				\m{g}^{\rm h}_2\\
			\end{bmatrix}
			\otimes
			\begin{bmatrix}
				\boldsymbol g^{\rm v}_1\\
				\boldsymbol g^{\rm v}_2\\
		\end{bmatrix}}
		{\boldsymbol{a}_{\rm Tx}^H}({\theta}_{\rm Tx}).
	\end{equation}
	
	If the number of subsurfaces is $K^2=1$, the piece-wise near-field channel model degenerates to the far-field case $\boldsymbol{G}_{\rm F}$ in (\ref{far-field}). If $K=N_{{\rm R}_y}=N_{{\rm R}_z}$, the piece-wise near-field model becomes the accurate traditional near-field channel model $\boldsymbol{G}_{\rm N}$ in (\ref{near-field}). In other words,  the piece-wise near-field channel model bridges the near-field and the far-field via fine tuning the number of subsurfaces. Comparing (\ref{near-field}), (\ref{far-field}), and (\ref{Mblock}), we can observe that the piece-wise channel model not only requires less number of parameters than the near-field model, but also enjoys a higher modeling accuracy than the far-field model. Indeed, the model presented in (9) is inspired by the model adopted in \cite{lu2023near}, which assumes the use of a ULA. Our proposed model in (9) considers the more practical uniform planar array (UPA) configuration of RIS, which is a more general model.	Moreover, inheriting from the near-field channel model, the piece-wise channel matrix in (\ref{Mblock}) still avails of a higher rank than the far-field channel matrix in (\ref{far-field}), i.e., $\rank(\m G_{\rm P})\geq \rank(\m G_{\rm F})=1$ \cite{liu2023near}, and thus it can improve the system's DoF and performance by exploiting the distance and angle diversity among different subsurfaces. In other words, the robustness stems from the fact that piece beamsteering is less prone to errors in distance and angle when there exist CEEs.

\section{Analysis of Interference Distribution for Different Channel Models}
In this section, we first introduce the CEE models and then analyze the distribution of interference-plus-noise signal for the three given models.
\subsection{Channel Estimation Error Models}
Based on different channel models, one can obtain the estimated channel via traditional training and parameter estimation procedures \cite{xie2016overview}, \cite{zheng2022survey}.  Note that when channel modeling is not accurate, the estimated channel suffers from not only CEEs, but also model mismatches. In particular, the actual channel between the Tx and RIS, which should follow a near-field channel model in the considered system, is composed of the estimated channel, the corresponding estimation error, and the model mismatch error. In the following, we represent the channel $\boldsymbol{G}_{\rm N}$ by different channel models:
	\begin{align}
		\boldsymbol{G}_{\rm N} &= \hat{\boldsymbol{G}}_{\rm N} + \Delta { \boldsymbol{G}}_{\rm N} +\Delta { \boldsymbol{M}}_{\rm N}, &\text{[Conventional]} \label{gn}\\
		\boldsymbol{G}_{\rm N} &= \hat{\boldsymbol{G}}_{\rm P} + \Delta { \boldsymbol{G}}_{\rm P} +\Delta { \boldsymbol{M}}_{\rm P}, \text{and} &\text{[Proposed]} \label{gp}\\
		\boldsymbol{G}_{\rm N} &= \hat{\boldsymbol{G}}_{\rm F} + \Delta { \boldsymbol{G}}_{\rm F} +\Delta { \boldsymbol{M}}_{\rm F}. &\text{[Far-field]} \label{gl}
	\end{align}
	For concise notation, we use subscripts $\{1\} = \{\rm N\}$, $\{2\} = \{\rm P\}$, and $\{3\} = \{\rm F\}$ to denote the conventional near-field, the proposed piece-wise near-field, and the far-field channel models, respectively, while $\hat{\boldsymbol{G}}_i$, $\Delta {\boldsymbol{G}}_i$, and $\Delta {\boldsymbol{M}}_i, i=1,2,3$, represent the estimated channel, the CEEs and the model mismatch error of different channel models, respectively. The left-hand side of (\ref{gn}), (\ref{gp}), and (\ref{gl}) is the ground-truth near-field channel state information $\boldsymbol{G}_{\rm N} $. As discussed in (\ref{near-field}), (\ref{far-field}), and (\ref{Mblock}), different channel  models represent $\boldsymbol{G}_{\rm N} $ with different channel matrix structures, i.e., $\boldsymbol{G}_{\rm N} = {\boldsymbol{G}}_i + \Delta {\boldsymbol{M}}_i, i=1,2,3 $. When $i=1$, this means that we adopt the near-field channel model for channel estimation which is free of model mismatch, i.e., $\Delta { \boldsymbol{M}}_1=\m 0$. Based on different channel models in (\ref{near-field}), (\ref{far-field}), and (\ref{Mblock}), channel estimation introduces additional CEE, i.e.,  $\boldsymbol{G}_i =\hat{\boldsymbol{G}}_i + \Delta {\boldsymbol{G}}_i, i=1,2,3 $.
	
	Assuming that the model mismatch errors are deterministic and the CEE follows a matrix-variate Gaussian distribution \cite{xing2009robust}, we have
	\begin{align}
	\boldsymbol G_{\rm N} &= \hat{\boldsymbol G}_i + \Delta {\boldsymbol{M}}_i + \Delta{\boldsymbol G}_i,  \ \text{and} \\		
	\Delta{\boldsymbol G}_i &\sim \mathcal{CN}_{N_{\rm R},N_{\rm Tx}}(\m 0, \sigma_{{\m G}_i}^2 \boldsymbol{\rm I}_{N_{\rm R}}\otimes \boldsymbol{\rm I}_{N_{\rm Tx}}).
\end{align}	
	Then, the distribution of the overall CSI imperfection	$\Delta \widetilde{\boldsymbol G}_i=\Delta { \boldsymbol{G}}_i +\Delta {\boldsymbol{M}}_i$ follows 
	\begin{align}
		\Delta \widetilde{\boldsymbol G}_i \sim \mathcal{CN}_{N_{\rm R},N_{\rm Tx}}(\Delta {\boldsymbol{M}}_i, \sigma_{{\m G}_i}^2 \boldsymbol{\rm I}_{N_{\rm R}}\otimes \boldsymbol{\rm I}_{N_{\rm Tx}}).
	\end{align}
	Similarly, the channel between the RIS and Rx is given by
	\begin{equation}
		\boldsymbol R = \hat{\boldsymbol R}+\Delta {\boldsymbol R},\Delta {\boldsymbol R} \sim \mathcal{CN}_{N_{\rm Rx},N_{\rm R}}(\boldsymbol 0, \sigma_{{\m R}}^2 \boldsymbol{\rm I}_{N_{\rm Rx}}\otimes \boldsymbol{\rm I}_{N_{\rm R}}),
	\end{equation}
	where $\hat{\boldsymbol R}$ and $\Delta {\boldsymbol R}$ represent the estimated channel and CEE of the RIS-Rx channel, respectively. To facilitate the subsequent analysis and design, we make the assumption that the channels of the Tx-RIS and RIS-Rx links are independently estimated and, thus, $\Delta \widetilde{\boldsymbol G}_i$ and $\Delta {\boldsymbol R} $  are independent of each other.\footnote{When dealing with a passive RIS, it becomes necessary to reconstruct the individual CSI of all links involving the RIS based on the acquired aggregate CSI, specifically the cascaded Tx-RIS-Rx CSI. This can be achieved by employing the methodology outlined in \cite{ardah2021trice}, \cite{zeng2022joint}.} 

\subsection{Covariance Matrix of the Interference-plus-Noise Signal}
For concise notation, let us drop the subscript for now. Based on the assumed CEE model of $\m{G}$ for the three channel models, we analyze the CEE distribution of the cascaded channel $\m{H}$ as below. Assuming that the estimations of the Tx-RIS and RIS-Rx channels are separately executed with their estimated values $\hat{\boldsymbol G}$ and $\hat{\boldsymbol R}$ \cite{wang2020channel}, \cite{zeng2022joint}, respectively, the estimated cascaded channel is given by
\begin{equation}
			\hat{\boldsymbol H} = \hat{\boldsymbol R} \boldsymbol \Phi \hat{\boldsymbol G}.
\end{equation}
Then, we define $\Delta\boldsymbol H$ and $\Delta{\boldsymbol H}_{\rm M}$ as the cascaded channel estimation error and the corresponding model mismatch error, respectively, which are given by
		\begin{align}
			\Delta\boldsymbol H &= \Delta\boldsymbol{R}\boldsymbol \Phi\hat{\boldsymbol G}+\hat{\boldsymbol R} \boldsymbol \Phi \Delta{\boldsymbol G}+\Delta\boldsymbol{R}\boldsymbol \Phi\Delta {\boldsymbol G}, \label{deltaH} \text{ and}\\
			\Delta\boldsymbol H_{\rm M} &= \hat{\boldsymbol{R}} \m{\Phi} \Delta \boldsymbol{M}+\Delta\boldsymbol{R} \m{\Phi} \Delta\boldsymbol{M}, \label{deltaHM}
		\end{align}
		respectively. The received signal at the Rx in (\ref{signalmodel}) can be reformulated as 
		\begin{equation}
			\boldsymbol y = (\hat{\boldsymbol H}+\Delta\boldsymbol H + \Delta\boldsymbol H_{\rm M} ) \boldsymbol{Ws} + \boldsymbol{n}=\hat{\boldsymbol H}\boldsymbol{Ws}+ \hat{\boldsymbol{n}},
		\end{equation}
		with the interference-plus-noise signal $\hat {\boldsymbol{n}} = (\Delta\boldsymbol H +\Delta\boldsymbol H_{\rm M})\boldsymbol{Ws}+ \boldsymbol{n}$.
		
		For the convenience of analysis, we assume that
		\begin{align}
			\|\Delta \m R\|_F &\ll \|\hat {\m R}\|_F,  \\
			\|\Delta \m G\|_F &\ll \|\hat {\m G}\|_F, \text{and} \\
			\|\Delta \m M\|_F &\ll \|\hat {\m G}\|_F,
		\end{align}
		respectively, which implies a relatively small CEE and model mismatch.\footnote{The assumptions are reasonable as the errors are significantly smaller than their estimated values, indicating a high level of confidence in the accuracy of the estimations. Referring to the CEE model in Section III-A, the estimated channel is significantly dominant, justifying this assumption.} By discarding the minor terms $\Delta\boldsymbol{R}\boldsymbol \Phi\Delta {\boldsymbol G}$ and $\Delta\boldsymbol{R} \m{\Phi} \Delta\boldsymbol{M}$ in (\ref{deltaH}) and (\ref{deltaHM}),  $\Delta \m H$ and $\Delta \m H_{\rm M}$ can be approximated as:
		\begin{align}
			\Delta\boldsymbol H &\approx \Delta\boldsymbol{R}\boldsymbol \Phi\hat{\boldsymbol G}+\hat{\boldsymbol R} \boldsymbol \Phi \Delta{\boldsymbol G}\text{ and}\\
			\Delta\boldsymbol H_{\rm M} &\approx \hat{\boldsymbol{R}} \m{\Phi} \Delta \boldsymbol{M},\label{deltaHMa}
		\end{align}
		respectively. Notice that $\Delta\boldsymbol H_{\rm M}$ in (\ref{deltaHMa}) is approximately deterministic when the passive beamforming strategy $\m \Phi$ and the estimated channel $\hat{\boldsymbol{R}}$ are given.

		The covariance matrix of the interference-plus-noise signal $\hat {\boldsymbol{n}}$ is given by
		\begin{equation}\label{modelerror1}
			\begin{aligned}
				 \boldsymbol{\Sigma}_{\hat {\boldsymbol{n}}} &= \mathbb{E}\{\hat {\boldsymbol{n}}\hat {\boldsymbol{n}}^{\it H}\} \\
				&= \Delta \boldsymbol{H}_{\rm M}{\boldsymbol W} {\boldsymbol W}^H \Delta \boldsymbol{H}^H_{\rm M} + \mathbb{E}\{\Delta \boldsymbol{H}{\boldsymbol W} {\boldsymbol W}^H \Delta \boldsymbol{H}^H \} +\sigma^2 {\boldsymbol {\rm I}}_{N_{\rm Rx}} \\
				&= \Delta \boldsymbol{H}_{\rm M}{\boldsymbol W} {\boldsymbol W}^H \Delta \boldsymbol{H}^H_{\rm M} 
				+\mathbb{E}\{\Delta\boldsymbol{R}\boldsymbol \Phi\hat{\boldsymbol G}{\boldsymbol W} {\boldsymbol W}^H \hat{\boldsymbol G}^H \boldsymbol\Phi^H \Delta\boldsymbol{R}^H\} \\
				  &+\mathbb{E}\{\hat{\boldsymbol R} \boldsymbol \Phi \Delta {\boldsymbol{G}} {\boldsymbol W} {\boldsymbol W}^H \Delta {\boldsymbol{G}}^H \boldsymbol\Phi^H \hat{\boldsymbol R}^H\} 
				+\sigma^2 {\boldsymbol {\rm I}}_{N_{\rm Rx}}.
			\end{aligned}
		\end{equation}	
		Next, we utilize the following lemma to calculate each term of $\boldsymbol{\Sigma}_{\hat {\boldsymbol{n}}}$ .
		\begin{lem} [\cite{zhang2017matrix}] \label{zhang}
			For a matrix $\boldsymbol{X}\in \mathbb{C}^{n \times m}$, which obeys the distribution
			$\boldsymbol{X} \sim {\mathcal{CN}_{n,m}}(\hat{\boldsymbol X}, \boldsymbol{R}_n\otimes\boldsymbol{R}_m) $, where 
			$\boldsymbol{R}_m\in \mathbb{C}^{m \times m}$ and $\boldsymbol{R}_n\in \mathbb{C}^{n \times n}$ represent the receive and transmit correlation matrices, respectively, and a compatible matrix $\boldsymbol{Z}$, it follows that
			\begin{center}
				$\mathbb{E}\left\{\boldsymbol X \boldsymbol Z {\boldsymbol X}^H\right\}=\hat{\boldsymbol X} \boldsymbol{Z} \hat{\boldsymbol X}^{H}+ \tr (\boldsymbol{Z}\boldsymbol{R}_m^T){\boldsymbol{R}_n}$.
			\end{center}
		\end{lem}
		
		It follows from Lemma \ref{zhang} that 
		\begin{equation}
		\begin{aligned}\label{oneerror}
				 \mathbb{E}\left\{\Delta\boldsymbol{R}\boldsymbol \Phi\hat{\boldsymbol G}{\boldsymbol W} {\boldsymbol W}^H \hat{\boldsymbol G}^H \boldsymbol\Phi^H \Delta\boldsymbol{R}^H \right\}
				 =  \sigma_{{\m R}}^2  \tr(\hat{\boldsymbol G} {\boldsymbol W} {\boldsymbol W}^H\hat{\boldsymbol G}^H) \boldsymbol{\rm I}_{N_{\rm Rx}}, \\
		\end{aligned}
		\end{equation}
		and
		\begin{equation}\label{twoerror}
			\begin{aligned}
				 \mathbb{E}\left\{\hat{\boldsymbol R} \boldsymbol \Phi \Delta {\boldsymbol{G}} {\boldsymbol W} {\boldsymbol W}^H \Delta {\boldsymbol{G}}^H \boldsymbol\Phi^H \hat{\boldsymbol R}^H\right\} 
				 = \sigma_{{\m G}}^2 \tr({\boldsymbol W} {\boldsymbol W}^H) \hat{\boldsymbol R} {\hat{\boldsymbol R}}^H.
			\end{aligned}		
		\end{equation}
		Inserting (\ref{oneerror}) and (\ref{twoerror}) into (\ref{modelerror1}), the covariance matrix of the interference-plus-noise signal in the presence of CEE and model mismatch is given by 
		\begin{equation}
			\begin{aligned}\label{ConxIpN}
				\boldsymbol{\Sigma}_{\hat {\boldsymbol{n}}} &= \Delta \boldsymbol{H}_{\rm M}{\boldsymbol W} {\boldsymbol W}^H \Delta \boldsymbol{H}^H_{\rm M}+\sigma_{{\m R}}^2 \tr(\hat{\boldsymbol G} {\boldsymbol W} {\boldsymbol W}^H\hat{\boldsymbol G}^H)\boldsymbol{\rm I}_{N_{\rm Rx}} 
				+\sigma_{{\m G}}^2\tr({\boldsymbol W} {\boldsymbol W}^H)\hat{\boldsymbol R} {\hat{\boldsymbol R}}^H +\sigma^2 {\boldsymbol {\rm I}}_{N_{\rm Rx}}.\\
			\end{aligned}
		\end{equation}	
		Substituting the subscript $i=\{1,2,3\}$, we can obtain the interference distributions for the three channel models caused by the CEEs and model mismatches, respectively.
		
		Note that for the near-field channel model, i.e., $\Delta \m M_1 = \m 0$, the covariance matrix $\boldsymbol{\Sigma}_{\hat {\boldsymbol{n}}}$ is influenced by the CSI error in the second and third terms and noise in the last term in (\ref{ConxIpN}), which is in line with the traditional imperfect CSI schemes that do not account for the model mismatch. On the other side, both the piece-wise near-field model and far-field model introduce additional interference due to the model mismatch in the first term in (\ref{ConxIpN}), which results in performance deterioration. Nevertheless, the piece-wise near-field model is a good compromise between the near-field and far-field models in terms of the number of channel model parameters and modeling accuracy. Indeed, the robustness of the proposed model arises from the system's DoF introduced by the near-field model, as well as the insensitivity of the far-field model to errors in distance and angle.

\section{Problem Formulation}
Due to the complicated variable coupling in the cascaded CEE when designing  $\m{W}$ and $\mathbf{\Phi}$ based on the estimated channel, we treat the interference caused by the model mismatch error and CEE as noise.\footnote{This represents a worst case assumption.} Then, the achievable SE between the transceivers is given by \cite{shi2011iteratively}
\begin{equation}\label{rate}
	{\cal R}(\m{W}, \m \Phi) =\log_2 \det (\boldsymbol{\rm I}_{N_{\rm Rx}}+{\hat{\boldsymbol H} {\boldsymbol W} {\boldsymbol W}^H\hat{\boldsymbol H}^H }{\boldsymbol{\Sigma}_{\hat{\boldsymbol{n}}}^{-1}}).
\end{equation}

Then, the joint active and passive beamforming design to maximize the achievable SE, under the transmit power constraint at the Tx and the constant modulus constraint for the phase control variable at the RIS, is formulated as the following optimization problem:
\begin{equation}\label{originalproblem}
	\begin{array}{ll}
		\max\limits_{ \boldsymbol{W}, \boldsymbol{\Phi}} & {\cal R} (\m{W}, \m \Phi)\\
		\text{s.t.} & \|\boldsymbol{W}\|^2_F\leq P_{\rm Tx},\\
		& |\m \phi_{n_{\rm R}}| =1, ~\forall {n_{\rm R}}=1, \cdots, N_{\rm R},  \\
	\end{array}
\end{equation}
where $\m \phi=\diag({\m \Phi})$. The constraint $\|\boldsymbol{W}\|^2_F\leq P_{\rm Tx}$ is introduced to prevent excessive power consumption and the constraint $|\m \phi_{n_{\rm R}}| =1$ is imposed on each diagonal entry of the phase control matrix $\m \Phi$ to maintain a constant modulus, which implies that the reflection coefficients applied by the RIS elements remain on the unit circle thereby simplifying the hardware implementation of the RIS.

Solving the SE maximization problem is generally challenging, particularly in the presence of CSI imperfections. The difficulty arises from the non-convex nature of the objective function ${\cal R} (\m{W}, \m \Phi)$, the non-convex constant modulus constraint $|\m \phi_{n_{\rm R}}| =1$, and even the intricate variable coupling between $\boldsymbol{W}$ and $ \boldsymbol{\Phi}$ as evident in the covariance matrix of the interference-plus-noise signal $\boldsymbol{\Sigma}_{\hat {\boldsymbol{n}}}$ in \eqref{ConxIpN}. Consequently, the optimization problem in (\ref{originalproblem}) is intractable and poses significant challenges for joint active and passive beamforming design.

By introducing two auxiliary variables $\boldsymbol{Z}\in \mathbb{C}^{N_{\rm Rx}\times N_{\rm s}}$ and $\boldsymbol\Omega  \in \mathbb{C}^{N_{\rm s}\times N_{\rm s}} \succeq \m 0$, the SE maximization problem in (\ref{originalproblem}) is equivalently transformed to an MSE minimization problem as follows \cite{zhao2023rethinking}:
	\begin{equation}\label{mmse}
		\begin{aligned}
			\left(\mathrm{P}\right)	\min\limits_{ \boldsymbol{W},  \boldsymbol{Z},  \boldsymbol{\Phi},\boldsymbol{\Omega}}  &\tr(\boldsymbol\Omega \boldsymbol{J}(\boldsymbol{Z, W}))-\log\det(\boldsymbol\Omega)-N_{\rm s} \\
			\text{ s.t. }\ & \|\boldsymbol{W}\|^2_F\leq P_{\rm Tx},\\
			& |\m \phi_{n_{\rm R}}| =1, ~\forall {n_{\rm R}}=1, \cdots, N_{\rm R},  \\
		\end{aligned}
	\end{equation}
	where
	\begin{align}\label{mse}
		\boldsymbol{J}(\boldsymbol{Z, W})&=\mathbb{E}\left\{(\boldsymbol{Z}^H\boldsymbol{y}-\boldsymbol{s})(\boldsymbol{Z}^H\boldsymbol{y}-\boldsymbol{s})^H\right\} \nonumber   \\
		&=\boldsymbol{Z}^H(\hat{\boldsymbol{H}}\boldsymbol{W}\boldsymbol{W}^H\hat{\boldsymbol{H}}^H +\boldsymbol{\Sigma}_{\hat{\boldsymbol n}})\boldsymbol{Z} 
		-\boldsymbol{Z}^H\hat{\boldsymbol{H}}\boldsymbol{W} 
		-\boldsymbol{W}^H\hat{\boldsymbol{H}}^H\boldsymbol{Z}+\boldsymbol{\rm I}_{N_{\rm s}},
	\end{align}
	is the MSE matrix function. The proof of the equivalence between the SE maximization problem in \eqref{originalproblem} and the MSE minimization problem in \eqref{mmse} follows a similar approach as \cite{shi2011iteratively, zhao2023rethinking}. In contrast to the SE maximization problem in \eqref{originalproblem}, the MSE minimization problem in \eqref{mmse} is convex regarding three variables, i.e., $\boldsymbol Z$, $\boldsymbol\Omega$, and  $ \boldsymbol{W}$, when $\boldsymbol{\Phi}$ is given. This observation paves the way for optimizing these four variables alternately via the BCD framework, as elaborated in the following section.

\section{Proposed Solution}
In the following, we introduce an iterative BCD approach to acquire an effective solution to (\ref{mmse}). The proposed approach divides (\ref{mmse}) into three subproblems that address different variables: 1) Optimize $\boldsymbol Z$ and $\boldsymbol\Omega$ given  $ \boldsymbol{W}$  and $\boldsymbol{\Phi}$; 2) Optimize $ \boldsymbol{W}$ given $\boldsymbol Z$, $\boldsymbol\Omega$ and $\boldsymbol{\Phi}$; 3) Optimize $\boldsymbol{\Phi}$ given $\boldsymbol Z$, $\boldsymbol\Omega$ and $ \boldsymbol{W}$. The following are the optimization steps for each of these three sub-problems.
\subsection{Update the Auxiliary Variables Matrices $\m Z$ and $\m \Omega$}
It is worth noting that when $\m{W}$ and $\m{\Phi}$ are given at each iteration, the optimal auxiliary variables $\boldsymbol{Z}$ and $\boldsymbol\Omega$ to minimize the objective function in (\ref{mmse}) are respectively given by
\begin{align}
	{\boldsymbol{Z}}&=(\hat{\boldsymbol{H}}\boldsymbol{W}\boldsymbol{W}^H\hat{\boldsymbol{H}}^H +\boldsymbol{\Sigma}_{\hat{\boldsymbol n}})^{-1}\hat{\boldsymbol{H}}\boldsymbol{W} \text{ and}  \label{zopt} \\ %
	{\boldsymbol{\Omega}}&=(\boldsymbol{J}({\boldsymbol Z}, \boldsymbol{W}))^{-1}.  \label{omegaopt}
\end{align}

\subsection{Update the Active Beamforming Matrix $\m W$}
%2.1) Precoding design
For given $\boldsymbol{Z}$, $\boldsymbol{\Omega}$, and $\boldsymbol{\Phi}$, the precoding matrix $\boldsymbol{W}$ can be updated by solving the following problem:
\begin{equation}
	\begin{aligned}
		\min \limits_{\boldsymbol{W}} &~ \tr(\boldsymbol{\Omega}(\boldsymbol{\rm I}_{N_{\rm s}}-\boldsymbol{Z}^H{\hat{\boldsymbol{H}}} \boldsymbol{W})(\boldsymbol{\rm I}_{N_{\rm s}}-\boldsymbol{Z}^H{\hat{\boldsymbol{H}}} \boldsymbol{W})^H)
		+\tr(\boldsymbol{\Omega}\boldsymbol{Z}^H \boldsymbol{\Sigma}_{\hat{\boldsymbol n}} \boldsymbol{Z})	\\
		\text{s.t.} &~  \|\boldsymbol{W}\|^2_F\leq P_{\rm Tx},
	\end{aligned}
\end{equation}
which is a convex optimization problem. Adopting the Lagrangian multiplier approach \cite{boyd2004convex}, the optimal active beamformer at the Tx is given by
\begin{align}\label{wopt}
	\boldsymbol{W} & = [{\hat{\boldsymbol{H}}}^H \boldsymbol{Z \Omega} \boldsymbol{Z}^H {\hat{\boldsymbol{H}}} + \sigma_{\m R}^2 \tr(\boldsymbol{\Omega} \boldsymbol{Z}^H\boldsymbol{Z}) {\hat{\boldsymbol{G}}}^H{\hat{\boldsymbol{G}}} \nonumber \\
	&+\sigma_{{\m G}}^2 \tr(\boldsymbol{\Omega} \boldsymbol{Z}^H  \hat{\boldsymbol R} {\hat{\boldsymbol R}}^H\boldsymbol{Z})\boldsymbol{\rm I}_{N_{\rm Tx}}+N_{\rm R} \sigma_{{\m G}}^2 \sigma_{{\m R}}^2 \tr(\boldsymbol{\Omega} \boldsymbol{Z}^H\boldsymbol{Z})\boldsymbol{\rm I}_{N_{\rm Tx}}
	+\eta \boldsymbol{\rm I}_{N_{\rm Tx}}]^{-1} {\hat{\boldsymbol{H}}}^H \boldsymbol{Z \Omega}, 
\end{align}
where the optimal Lagrangian multiplier  $\eta \geq 0$  can be found by the proposed algorithm in \cite{pan2020multicell}. Note that to maximize the achievable SE, we have $\|\boldsymbol{W}\|^2_F= P_{\rm Tx}$ at the optimum \cite{zeng2022joint}.

\subsection{Update the Passive Beamforming Matrix $\m \Phi$ } 
% 2.2) ADPM-based phase control 
For given $\boldsymbol{Z}$, $\boldsymbol{\Omega}$, and $\boldsymbol{W}$, the passive beamforming design problem is formulated as
\begin{align} \label{sub2}
	\begin{aligned}
		\min\limits_{\boldsymbol{\Phi}} \ & f(\boldsymbol{\Phi}) =\tr(\boldsymbol{\Omega}(\boldsymbol{\rm I}_{N_{\rm s}}-\boldsymbol{Z}^H{\hat{\boldsymbol{H}}} \boldsymbol{W})(\boldsymbol{\rm I}_{N_{\rm s}}-\boldsymbol{Z}^H{\hat{\boldsymbol{H}}} \boldsymbol{W})^H) 
		+ \ \tr(\boldsymbol{\Omega}\boldsymbol{Z}^H \boldsymbol{\Sigma}_{\hat{\boldsymbol n}} \boldsymbol{Z})\\
		\text{s.t.} \ & |\m \phi_{n_{\rm R}}|=1,  ~\forall {n_{\rm R}}=1, \ldots, N_{\rm R}.  \\
	\end{aligned}
\end{align}
Substituting $\hat{\boldsymbol{H}}$ into (\ref{sub2}) and ignoring items that are not related to $\boldsymbol{\Phi}$, the objective function in (\ref{sub2}) can be simplified as
\begin{align}
	f(\boldsymbol{\Phi}) & = \tr({\boldsymbol{\Phi}}^H {\hat{\boldsymbol{R}}}^H {\boldsymbol{Z \Omega}} \boldsymbol{Z}^H {\hat{\boldsymbol{R}}} {\boldsymbol{\Phi}}  {\hat{\boldsymbol{G}}} \boldsymbol{W} \boldsymbol{W}^H {\hat{\boldsymbol{G}}}^H)  -\tr({\hat{\boldsymbol{R}}}^H {\boldsymbol{Z \Omega}}\boldsymbol{W}^H{\hat{\boldsymbol{G}}}^H{\boldsymbol{\Phi}}^H)- \tr({\hat{\boldsymbol{G}}}\boldsymbol{W}{\boldsymbol{\Omega}}\boldsymbol{Z}^H{\hat{\boldsymbol{R}}}{\boldsymbol{\Phi}}). 
\end{align}
According to \cite[Lemma 10.6]{zhang2017matrix},  we can further simplify the optimization problem in (\ref{sub2}) to a standard quadratic programming (QP) problem, i.e.,
\begin{align} \label{fphi}
	\begin{array}{ll}
		\min\limits_{\boldsymbol{\Phi}} &~ f(\boldsymbol\phi) = {\boldsymbol\phi} ^H \boldsymbol{A}{\boldsymbol\phi}-2\Re\{\boldsymbol{d}^T {\boldsymbol \phi} \}\\
		\text{s.t.} &  |\m \phi_{n_{\rm R}}|=1,  ~\forall {n_{\rm R}}=1, \ldots, N_{\rm R},
	\end{array} 
\end{align} 
where
\begin{align}
	\boldsymbol{A} &=({\hat{\boldsymbol{R}}}^H {\boldsymbol{Z \Omega}} \boldsymbol{Z}^H {\hat{\boldsymbol{R}}})  \circ  ({\hat{\boldsymbol{G}}} \boldsymbol{W} \boldsymbol{W}^H {\hat{\boldsymbol{G}}}^H)^T,  \label{A} \\
	\boldsymbol{d} &= (\boldsymbol{D}_{1, 1},\dots,\boldsymbol{D}_{N_{\rm R}, N_{\rm R}})^T, \text{ and }  \boldsymbol{D}={\hat{\boldsymbol{G}}}\boldsymbol{W}{\boldsymbol{\Omega}}\boldsymbol{Z}^H{\hat{\boldsymbol{R}}}. \label{d} 
\end{align} 

We notice that the constant modulus constraint in equation (\ref{fphi}) is generally non-convex and NP-hard, posing a challenge for solving the quadratic optimization problem. In contrast to the traditional alternating direction method of multipliers (ADMM) framework, which employs a fixed penalty factor, our approach is inspired by the method of multipliers and the penalty alternating direction methods discussed in \cite{ghadimi2014optimal}, \cite{magnusson2015convergence} to introduce an alternating direction penalty method (ADPM) algorithm. This algorithm gradually increases the penalty factor during iterations to drive the penalty term toward zero, thereby facilitating the design of the phase of the RIS. More specifically, by introducing an auxiliary variable $\boldsymbol\phi_0 = \left[e^{j\vartheta_1}, \ldots, e^{j\vartheta_{N_{\rm R}}} \right] \in \mathbb{C}^{N_{\rm R} \times 1}$, we equivalently recast the problem in (\ref{fphi}) as
\begin{equation}\label{goal7}
	\begin{aligned}
		\min \limits_{\boldsymbol \phi, \boldsymbol \phi_0} &~ {\boldsymbol\phi} ^H \boldsymbol{A}{\boldsymbol\phi}-2\Re\{\boldsymbol{d}^T {\boldsymbol \phi} \} \\
		\text { s.t. } & \boldsymbol \phi=\boldsymbol\phi_0,\\
		&  |{\boldsymbol\phi_0}_{n_{\rm R}}| =1, ~\vartheta_{n_{\rm R}} \in [0,2\pi], ~\forall {n_{\rm R}}=1, \ldots, N_{\rm R}.  \\
	\end{aligned}
\end{equation}
The augmented Lagrangian function of (\ref{goal7}) is given by
\begin{align}\label{alf}
	{\cal L} ={\it f}(\boldsymbol\phi)+\Re\{{\cal \boldsymbol u}^{\it H} (\boldsymbol\phi-\boldsymbol\phi_0)\}+\frac{\rho}{2}||\boldsymbol\phi-\boldsymbol\phi_0||_2^2, 
\end{align}
where ${\it f}(\boldsymbol\phi)={\boldsymbol\phi} ^H \boldsymbol{A}{\boldsymbol\phi}-2\Re\{\boldsymbol{d}^T {\boldsymbol \phi} \}$, while ${\cal{\boldsymbol u}} \in \mathbb{C}^{N_{\rm R} \times 1}$ and  $\rho > 0$ are the multiplier vector and the penalty factor, respectively.

In the following, we illustrate how to update $\boldsymbol\phi_0$ and $\boldsymbol\phi$, and then discuss the selection of $\rho^{(0)}$.
\subsubsection{Update $\boldsymbol\phi_0$}
When we consider the update of $\boldsymbol\phi_0$ given $\boldsymbol\phi^{(t-1)}, {\cal\boldsymbol u}^{(t-1)}, \rho^{(t-1)}$ in the $t$-th iteration of ADPM, we omit the constant terms in ${\cal L}$ that are irrelevant to $\boldsymbol\phi_0$, and the optimization problem is given by
\begin{equation}\label{phi0}
	\begin{array}{ll}
		\min\limits_{\boldsymbol\phi_0}& \Re\{(-{\cal\boldsymbol u}^{(t-1)}-\rho^{(t-1)}\boldsymbol\phi^{(t-1)})^H \boldsymbol\phi_0 \}\\
		\text{s.t.} & |\boldsymbol\phi_0|=1, ~ \vartheta_{n_{\rm R}} \in [0,2\pi], ~n_{\rm R} = 1,\ldots, N_{\rm R}.
	\end{array}
\end{equation}
The optimal solution of problem (\ref{phi0}) is given as 
\begin{equation}\label{optphi0}
	\vartheta_{n_{\rm R}} = \angle \{\m\gamma^{(t-1)}_{n_{\rm R}}\},
\end{equation}
where $\boldsymbol\gamma^{(t-1)}={\cal\boldsymbol u}^{(t-1)}+\rho^{(t-1)}\boldsymbol\phi^{(t-1)} \in \mathbb{C}^{N_{\rm R} \times 1}$. 

\subsubsection{Update  $\boldsymbol\phi$}
When we consider the update of $\boldsymbol\phi$ given $\boldsymbol\phi_0^{(t)}, {\cal\boldsymbol u}^{(t-1)}, \rho^{(t-1)}$, the minimization problem is given by
\begin{equation}\label{phi}
	\min\limits_{\boldsymbol\phi} f(\boldsymbol\phi) + \Re\{({\cal\boldsymbol u}^{(t-1)}-\rho^{(t-1)}\boldsymbol\phi_0^{(t)})^H \boldsymbol\phi\}.
\end{equation}
We can obtain the closed-form optimal solution  to the problem in (\ref{phi}) as
\begin{equation}\label{optphi}
	\boldsymbol\phi^{(t)} = (2{\boldsymbol A}+{\rho^{(t-1)}} \boldsymbol {\rm I} )^{-1}(\rho^{(t-1)}\boldsymbol\phi_0^{(t)}-{\cal\boldsymbol u}^{(t-1)}+2\boldsymbol{d}).
\end{equation}

\begin{algorithm}[t]  \label{ADPM}
	\caption{ADPM-based Algorithm for Handling (\ref{goal7})}
	\label{alg:Framwork}
	\begin{algorithmic}[1]
		\STATE {\textbf{Initialize:} $\boldsymbol{A}, \boldsymbol{d}, {\cal\boldsymbol u}^{(0)}, \rho^{(0)}>0, \delta_1, \delta_2, \epsilon, \kappa$, where $0<\delta_1<1$, $\delta_2>1$ are close to 1.}
		\WHILE{$\Delta e^{(t)}=||\boldsymbol\phi^{(t)}-\boldsymbol\phi_0^{(t)}||> \epsilon$} 
		\STATE {\textbf{update:} $\boldsymbol\phi_0^{(t)}$ and $\boldsymbol\phi^{(t)}$
			\begin{equation*}
				\angle \{{(\m \phi^{(t)}_0)} _{n_{\rm R}}\} = \angle \{\m\gamma^{(t-1)}_{n_{\rm R}}\},
			\end{equation*}
			\begin{equation*}
				\boldsymbol\phi^{(t)} = (2{\boldsymbol A}+{\rho^{(t-1)}} \boldsymbol {\rm I}_{N_{\rm R} })^{-1}(\rho^{(t-1)}\boldsymbol\phi_0^{(t)}-{\cal\boldsymbol u}^{(t-1)}+2\boldsymbol{d}),
		\end{equation*}}
		\STATE {\textbf{update:} ${\cal\boldsymbol u}^{(t)}$ and $\rho^{(t)}$
			\begin{align*}\label{third}
				\begin{split}
					\rho^{(t)}=\left\{
					\begin{array}{ll}
						\rho^{(t-1)}, &\Delta e^{(t)}\leq \delta_1 \Delta e^{(t-1)},\\
						\delta_2 \rho^{(t-1)}, & \text{else}.
					\end{array}
					\right.
				\end{split}
			\end{align*}
			\begin{equation*}\label{forth}
				\begin{split}
					{\cal\boldsymbol u}^{(t)}=\left\{
					\begin{array}{ll}
						{\cal\boldsymbol u}^{(t-1)}+\rho^{(t)}(\boldsymbol\phi^{(t)}-\boldsymbol\phi_0^{(t)}), \ u^{(t)}_{\max}\leq \kappa,\\
						({\cal\boldsymbol u}^{(t-1)}+\rho^{(t)}(\boldsymbol\phi^{(t)}-\boldsymbol\phi_0^{(t)}))/u^{(t)}_{\max}, \ \text{else},
					\end{array}
					\right.
				\end{split}
			\end{equation*}
			where $u^{(t)}_{\max}$ is the value with the largest modulus in the multiplier vector ${\cal\boldsymbol u}^{(t)}$.}
		\ENDWHILE
		\STATE {\textbf{Output:} $\boldsymbol{\phi}^{\ast}$}   
	\end{algorithmic}
\end{algorithm}

The ADPM-based algorithm for designing the passive beamforming at the RIS is outlined in Algorithm 1, detailing the update rules for $\rho$ and $\cal\boldsymbol u$.  If we substitute the update rules of $\rho^{(t)}$ and ${\cal\boldsymbol u}^{(t)}$ in Algorithm 1 with $\rho^{(t)}=\rho^{(t-1)}$ and ${\cal\boldsymbol u}^{(t)}={\cal\boldsymbol u}^{(t-1)}+\rho^{(t)}(\boldsymbol\phi^{(t)}-\boldsymbol\phi_0^{(t)})$, Algorithm 1 degenerates to the classical ADMM framework. According to the theoretical analysis in \cite{yu2020quadratic}, adapting the penalty factor $\rho^{(t)}$ is crucial with the following principle: increasing it when the primal residual $\Delta e^{(t)}$ fails to decrease with iterations, helps drive $\Delta e^{(t)}$ towards zero to locate a feasible point. Otherwise, $\rho^{(t)}$ remains unchanged. This strategy aims to enhance the likelihood of the ADPM algorithm discovering a feasible point compared to the ADMM \cite{yu2020quadratic}, \cite{yu2020discrete}. 

Moreover, achieving faster convergence can be facilitated by selecting an appropriate initial value for the penalty factor $\rho^{(0)}$. In this regard, we employ a method proposed in \cite{ghadimi2014optimal} to determine the initialized penalty factor for the ADPM. Specifically, when problem (\ref{goal7}) does not involve any non-convex constraints, an effective initialized penalty factor for ADPM is obtained as \cite{ghadimi2014optimal}:
\begin{equation}
	\rho^{(0)} = \sqrt{\lambda_{\min}(\m A) \lambda_{\max}(\m A)},
\end{equation}
where $\lambda_{\min}(\m A)$ and $\lambda_{\max}(\m A)$ represent the minimum and maximum eigenvalues of $\m A$, respectively. It should be noted that if the smallest eigenvalue of $\m A$ is zero, $\lambda_{\min}(\m A)$ is assigned to its smallest nonzero eigenvalue. For further details, please refer to Theorem 4 in \cite{ghadimi2014optimal}.

\begin{Remark}
	The proposed ADPM algorithm is guaranteed to converge for arbitrary initialization $\boldsymbol\phi^{(0)}$ and ${\cal\boldsymbol u}^{(0)}$ provided that
	$\rho^{(0)}>0, 0<\delta_1<1, \delta_2>1$, and $\kappa$ is a sufficiently large positive number \cite{yu2020quadratic}. In particular, for the continuous phase case, if the penalty
	parameter $\rho^{(t)}$ is bounded, the limiting point $\boldsymbol\phi^{\ast}$ of the sequence $\{\boldsymbol\phi^{(t)}\}_{t=0}^\infty$ obtained via the ADPM algorithm is a Karush-Kuhn-Tucker (KKT) point
	of (\ref{goal7}) \cite{yu2020quadratic}, \cite{yu2020discrete}. 
\end{Remark}

\begin{algorithm}[t]
	\caption{Overall Block Coordinate Descent Algorithm for Addressing (\ref{mmse})}
	\label{alg:MSE-ADPM}
	\begin{algorithmic}[1]	
		\STATE {\textbf{Initialize:} $\boldsymbol{\Phi}^0$, $\boldsymbol{W}^0$, $\boldsymbol{Z}^{0}$ and $\boldsymbol{\Omega}^{0}$, tolerance accuracy $\varepsilon$, maximum number of iterations $r_{\max}$, the objective function value of problem (\ref{mmse}) $P(\boldsymbol{\Phi}^0$, $\boldsymbol{W}^0)$.}	 \\
		\REPEAT 
		\STATE Given $\boldsymbol{W}^{r}, \boldsymbol{\Phi}^{r}$ and $\boldsymbol{\Omega}^{r}$, compute the auxiliary variable $\boldsymbol{Z}^{r}$ by (\ref{zopt}).\\ 
		\STATE Given $\boldsymbol{W}^{r}, \boldsymbol{\Phi}^{r}$ and $\boldsymbol{Z}^{r}$, compute the auxiliary variable $\boldsymbol{\Omega}^{r}$ by (\ref{omegaopt}).\\
		\STATE Given $\boldsymbol{Z}^{r}$, $\boldsymbol{\Omega}^{r}$ and $\boldsymbol{\Phi}^{r}$, determine $\eta$ and compute $\boldsymbol{W}^{r+1}$ by (\ref{wopt}). \\
		\STATE Given $\boldsymbol{Z}^{r}$, $\boldsymbol{\Omega}^{r}$ and $\boldsymbol{W}^{r+1}$, compute $\boldsymbol{A}$ and $\boldsymbol{d}$ by (\ref{A}) and (\ref{d}) .\\
		\STATE Update $\boldsymbol{\phi}^{r+1}$ by Algorithm 1, and reconstruct $\boldsymbol{\Phi}^{r+1}$.\\
		\STATE Set $r=r+1$.\\
		\UNTIL{ $r>r_{\max}$ or $|P(\boldsymbol{W}^{r+1},\boldsymbol{\Phi}^{r+1})-P(\boldsymbol{W}^{r},\boldsymbol{\Phi}^{r})|<\varepsilon$} \\
		\STATE {\textbf{Output:} $\boldsymbol{\Phi}^{\ast}, \boldsymbol{W}^{\ast}$.}\\
	\end{algorithmic}
\end{algorithm}

\subsection{Overall Algorithm and Complexity Analysis}
Now, we provide the detailed description of the overall BCD algorithm for solving (\ref{mmse}) in Algorithm 2. Step 1 is used to initialize these variables and to set thresholds. Steps 2 and 3 are used to update the auxiliary variables $\m Z$ and $\m \Omega$, step 4 updates the active beamforming matrix at the Tx. Then, steps 5 and 6 update the passive beamforming matrix of the RIS by ADPM proposed in Algorithm 1. Finally, the stopping condition is $|P(\boldsymbol{W}^{r+1},\boldsymbol{\Phi}^{r+1})-P(\boldsymbol{W}^{r},\boldsymbol{\Phi}^{r})|<\varepsilon$. The convergence analysis of the proposed Algorithm 2 can be found in \cite{pan2020multicell}, \cite{zeng2022joint}. 

Note that the proposed problem formulation and algorithmic solution are applicable for all the three channel models in (\ref{near-field}), (\ref{far-field}), and (\ref{Mblock}). However, the impact of channel models on the system performance are characterized by the covariance matrices $\boldsymbol{\Sigma}_{\hat {\boldsymbol{n}}}$ in the problem formulation (\ref{rate}).
A further extension of this work is to leverage the matrix structures of $\boldsymbol{\Sigma}_{\hat {\boldsymbol{n}}}$ and $\hat{\m H}$ in different channel models to design specific algorithms aimed at exploring the impact of channel models on the system performance for RIS-aided MIMO communications.

 We note that the original problem is divided into three sub-problems and addressed iteratively, which requires $I_{\rm O}$ iterations. For the update of the two auxiliary variables, $\{\boldsymbol{Z}\}$, $\{\boldsymbol{\Omega}\}$, it requires the computation of order $\mathcal{O}(N^3_{\rm Rx})$ and $\mathcal{O}(N^3_{\rm s})$, respectively. For the active beamforming problem at the Tx,  solving $\{\boldsymbol{W}\}$  requires the computation of order $\mathcal{O}(I_{\eta} N^3_{\rm Tx})$, where $I_{\eta}$ is the number of iterations for searching the dual variable $\eta$. Since the number of transmitter antennas $N_{\rm Tx}$ is usually larger than $N_{\rm s}$ and $N_{\rm Rx}$, the complexity of the second sub-problem is $\mathcal{O}(I_WI_{\eta}N^3_{\rm Tx})$, where $I_W$ is the number of iterations required to converge. For the passive beamforming problem at the RIS, the complexity of the third sub-problem is $\mathcal{O}(N^3_{\rm R}+I_A N^2_{\rm R})$, where $I_A$ is the number of iterations required to converge. Based on the above analysis, the computational complexity of Algorithm \ref{alg:MSE-ADPM} is $\mathcal{O}(I_{\rm O}(N^3_{\rm Rx}+N^3_{\rm s}+I_WI_{\eta}N^3_{\rm Tx}+N^3_{\rm R}+I_A N^2_{\rm R}))$.

\section{Numerical Results}
In this section, we present simulation results to assess the performance of the three different channel models in the presence of channel estimation error. A 3D Cartesian coordinate system is considered, where the BS, the RIS, and the Rx are located at $(10, -20, 5)$ m, $(0, 0, 10)$ m, and $(100, 50, 5)$ m, respectively. The Tx is equipped with $N_{\rm Tx}=64$ transmit antennas serving one user equipment (UE) equipped with $N_{\rm Rx}=8$ receive antennas with the assistance of an RIS. The number of data streams is $N_{\rm s}=64$. The number of reflection elements is $N_{\rm R}= 256$. The path loss model utilized is a model tailored for RIS-aided near-field communication, as detailed in \cite{bjornson2020power}. The carrier frequency is $30$ GHz and the number of Monte Carlo experiments is $50$. The ground-truth channel from the BS to the RIS follows a near-field channel model while the system design adopts three different channel models: the conventional near-field, the proposed piece-wise near-field and the far-field channel models, resulting in different covariance matrices of the interference-plus-noise signal. Meanwhile, the channel from the RIS to the Rx follows a far-field channel model. The transmit SNR is defined by $\text{SNR} = 10\log_{10}(P_{\rm Tx}/\sigma^2)$, where $P_{\rm Tx}$ and $\sigma^2$ is the power of the transmit signal and noise, respectively. Unless otherwise specified, we set $ {\sigma^2} = -80$ dBm. 
	
 The initialization parameters of the ADPM-based algorithm for solving (\ref{goal7}) and Algorithm 2 are provided as follows: the multiplier vector ${\m u} ^{(0)} = \m 0$, $\epsilon = 10^{-6}$, $\delta_1 = 0.95$, $\delta_2 = 1.05$, $\kappa = 10^{3}$, $\varepsilon = 10^{-3}$, and $r_{\max} = 100$. For more detailed parameter settings, please refer to the reference \cite{yu2020quadratic}. 
We assume an identical normalized CEE for different channel models, i.e., $\sigma^2_{{\m G}_i}= \tau_i \cdot \mathbb{E}\{\|\m G_{\rm N}- \Delta \m M_i\|^2_F\}$, where $\tau_i$ is the normalized CEE for the Tx-RIS link and it is given by
\begin{equation}
	\tau_{i} = \frac{\mathbb{E}[||\boldsymbol{G}_{\rm N}- \Delta{\boldsymbol{M}_i}-\hat {\boldsymbol{G}_i}||^2_F]}{\mathbb{E}[||\boldsymbol{G}_{\rm N} - \Delta{\boldsymbol{M}_i}||^2_F]}, i=1, 2, 3.
\end{equation}

Similarly, we can define the normalized CEE $\tau_{\m R}$  for the RIS-Rx link and it is given by
\begin{equation}
	\tau_{\m R} = \frac{\mathbb{E}[|| \Delta{\boldsymbol{R}}||^2_F]}{\mathbb{E}[||\boldsymbol{R}||^2_F]}.
\end{equation}

\subsection{Convergence Validation} 
%\vspace{-10pt}
\begin{figure}[t]
	\centering
	\includegraphics[scale=0.6]{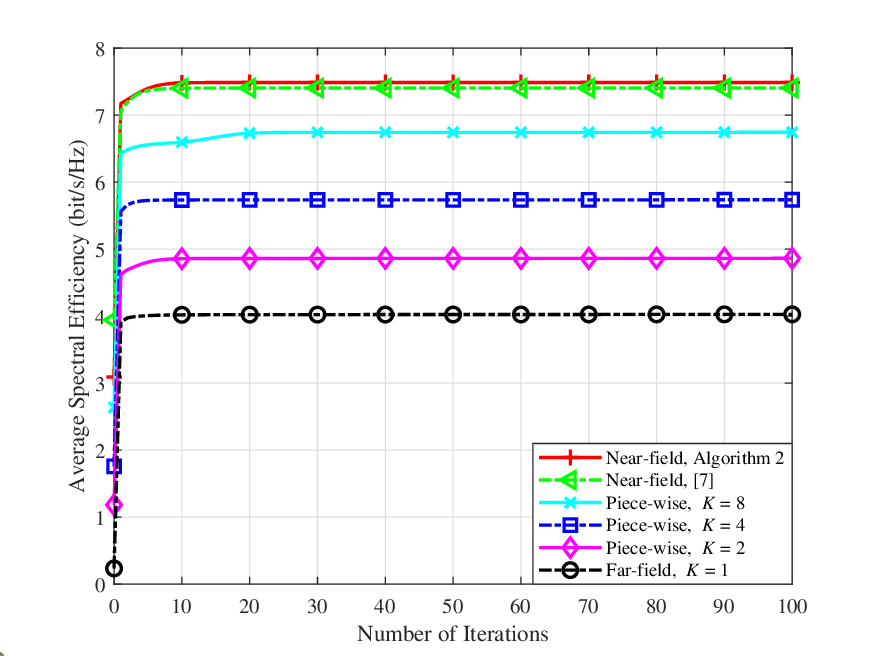}
	\caption{ Convergence behavior when $d_{\rm BR}=20 $ m and SNR = 10 dB.}
	\label{fig: convergence}
\end{figure}
We investigate the average achievable SEs for different channel models without considering any estimation error in Fig. \ref{fig: convergence}, i.e., $\tau_i=0, i=1, 2, 3$, which means that only model mismatch errors are considered. Here, we set the distance along the y-axis from the Tx to the RIS as $d_{\rm BR}=20$ m and ${\rm SNR} = 10$ dB. We compare the performance of our proposed algorithm (i.e., Algorithm 2), with that of the algorithm in \cite{zhang2020capacity} under the near-field channel model, verifying the effectiveness and convergence of Algorithm 2. When $K=1$, i.e., the RIS is not partitioned into subsurfaces, the piece-wise near-field channel model degenerates to the conventional far-field channel model. It can be seen that the performance of the piece-wise near-field channel model with multiple subsurface structures is indeed better than that of the traditional far-field model, owing to the reduced model mismatch error as well as the increased DoF. Furthermore, as the number of subsurface increases, adopting the piece-wise near-field channel model gradually approaches the performance of the conventional near-field channel model, where the latter model does not have any model mismatch error, which is consistent with the channel model analysis in Section II-B. It is worth noting that in the extreme case, where the RIS is divided into 256 subsurfaces, the piece-wise near-field channel model evolves into the near-field channel model. Therefore, dividing the RIS into 64 pieces of subsurfaces, i.e., $K = 8$, can achieve the dominant performance gain of the conventional near-field model, while this piece-wise near-field channel model significantly reduces the number of parameters involved.

\subsection{SE vs SNR for Different Channel Models in the Presence of CEE } 
\begin{figure}[t]
	\centering
	\includegraphics[scale=0.6]{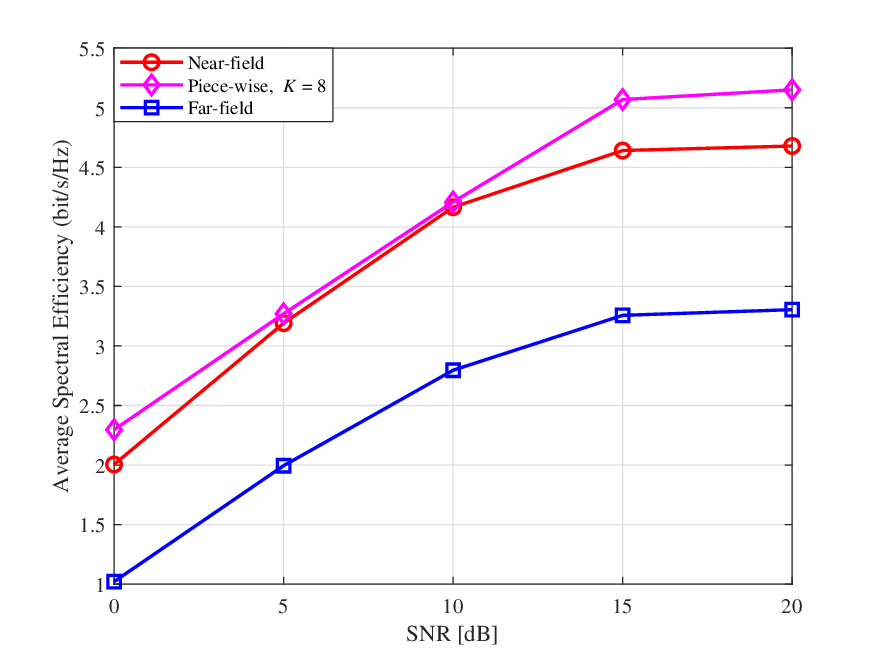}
	\caption{The achievable SEs versus the SNR when $\tau = 0.2$.}
	\label{fig: SNR}
\end{figure}	
Figure \ref{fig: SNR} presents the average achievable SEs for different channel models at different SNR levels in the presence of CEE with $\tau=0.2$.\ The slope of the SE curve corresponding to each model represents the multiplexing gain with a steeper model indicating more available DoF. The results depict that the proposed piece-wise near-field channel provides more DoF compared to the traditional far-field channel, potentially leading to an increased achievable SE. Although the DoF provided by the near-field channel model are slightly higher than that of the piece-wise near-field channel model, adopting the piece-wise channel model can achieve a higher achievable SE than that of the conventional near-field model, due to the high sensitivity of the latter model to CEEs. This is because beamsteering is robust against the beam misalignment due to the angle and distance errors, and it exploits more DoF brought by the conventional near-field channel model. Besides, we note that the achievable SEs for all the three channel models are saturated in the high SNR regime due to the presence of CEEs and potential model mismatches.

\subsection{SE vs Normalized CEE for Different Channel Models} 
\begin{figure}[t]
	\centering
	\includegraphics [scale=0.6]{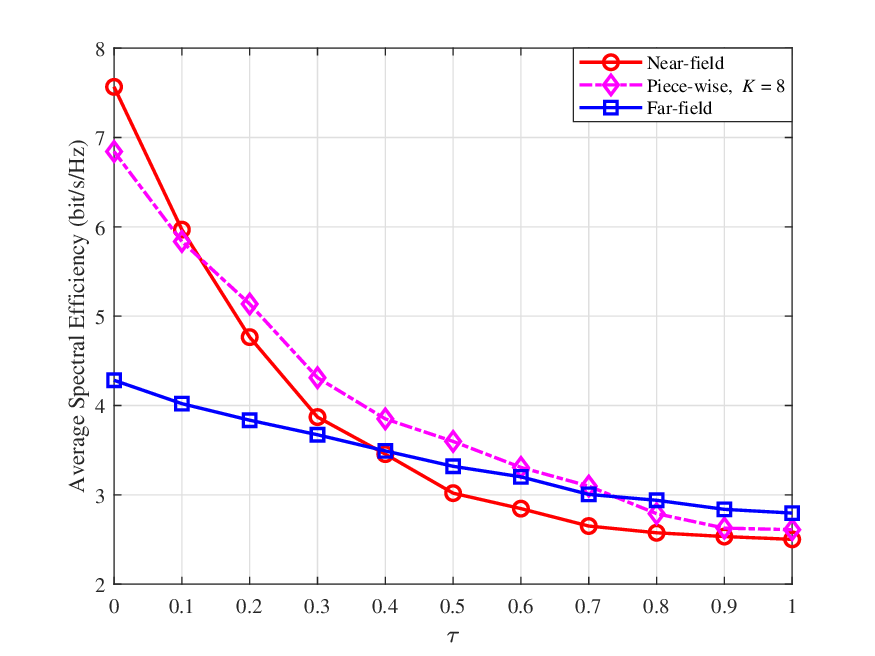}
	\caption{ The achievable SEs versus the CEE variance when $K = 8$.}
	\label{fig: CEE}
\end{figure}
In Fig. \ref{fig: CEE}, the average achievable SEs for different channel models are presented under different  normalized CEEs for $K=8$. When the normalized CEE $\tau$ is 0, it corresponds to the scenario without any estimation error in Fig. 2. The results show that as the normalized CEE $\tau$ increases, the performance of the near-field channel model deteriorates significantly, which indicates the high sensitivity of beamfocusing with respect to CEE in the near-field region. We observe that when the normalized CEE $\tau>0.13$, the proposed piece-wise near-field model yields a better performance than the near-field model due to the enhanced robustness inherited from the far-field model. When the normalized CEE is large, the performances of the piece-wise near-field and near-field models are nearly identical as the beamfocusing in both cases is inaccurate. When the normalized CEE $\tau>0.75$, the performance of the far-field channel model is better than that of both the near-field and piece-wise near-field models due to its robustness to CEE. This implies that for different levels of CEE, different channel models can be adopted to improve the system performance. By combining the results from Fig. \ref{fig: SNR} and Fig. \ref{fig: CEE}, it becomes evident that the piece-wise channel model not only surpasses the DoF associated with the far-field model but also enhances the system robustness against the CEEs when compared to the near-field model.

\subsection{SE vs Number of Transmit Antennas for Different Channel Models}
\begin{figure}[htbp]
	\centering
	\includegraphics [scale=0.6]{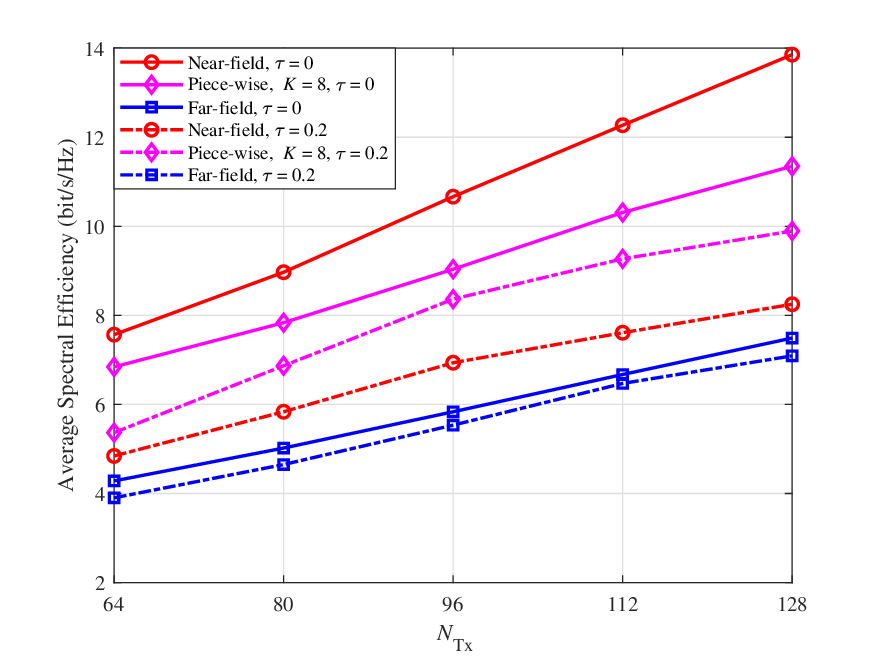}
	\caption{ The achievable SEs versus the number of transmit antennas, $N_{\rm {Tx}}$, when $N_{\rm R}=256$.}
	\label{fig: NTX}
\end{figure}
Figure \ref{fig: NTX} demonstrates the impact of the number of transmit antennas at the Tx on the achievable SE. We present the achievable SEs for three models under two scenarios: perfect CSI ($\tau=0$) and imperfect CSI ($\tau=0.2$).  When $\tau=0$, a linear scaling in the SE is observed with respect to the number of transmit antennas. In this scenario, the deterministic model mismatch $\Delta \m H_{\rm M}$ in (\ref{ConxIpN}) is considered as interference, and increasing $N_{\rm Tx}$ provides more DoF for designing the active beamforming matrix $\m W$ to suppress the interference caused by the model mismatch, i.e., the first term in (\ref{ConxIpN}), thus approaching interference-free transmission. However, when $\tau=0.2$, the DoF are not sufficient to design the active beamforming matrix to suppress the interference caused by the uncertain CEE, i.e., the second and third term in (\ref{ConxIpN}). Moreover, the piece-wise near-field model with imperfect CSI demonstrates a higher SE than the far-field model with perfect CSI, highlighting the DoF advantages introduced by the piece-wise near-field model. 

\subsection{SE vs Number of Reflecting Elements for Different Channel Models} 
\begin{figure}[htbp]
	\centering
	\includegraphics [scale=0.6]{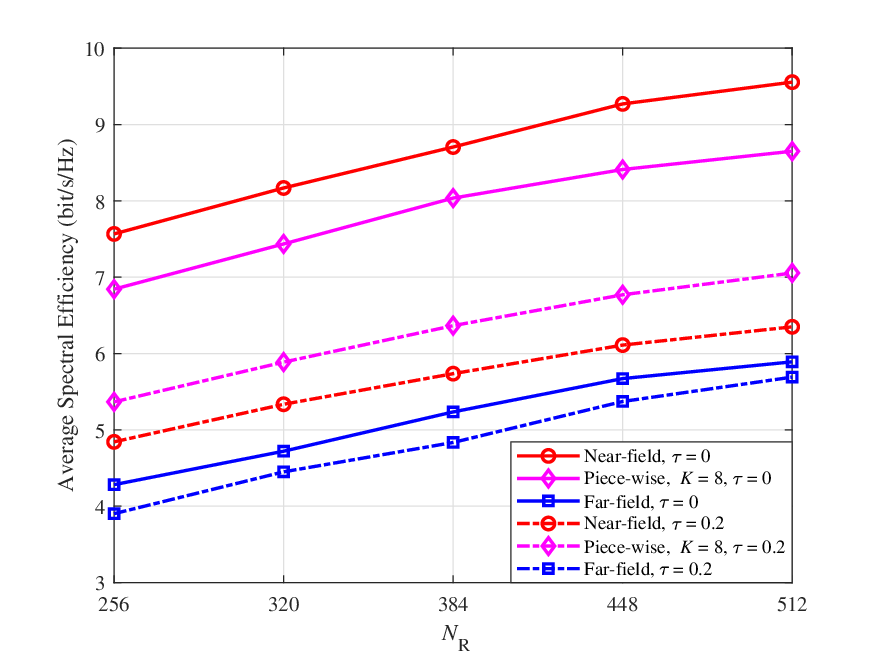}
	\caption{ The achievable SEs versus the number of reflecting elements $N_{\rm R}$ when $N_{\rm {Tx}}=64$.}
	\label{fig: RIS}
\end{figure}
In Fig. \ref{fig: RIS}, we investigate the impact of the number of reflecting elements on the achievable SEs. We present SEs for the three models with varying numbers of reflecting elements under perfect CSI ($\tau=0$) and imperfect CSI ($\tau=0.2$) scenarios. We observe that as the number of reflecting elements increases, the SE of the three models exhibits linear growth when $N_{\rm R}$ is small. However, the slope of the SE curves tends to flatten for larger numbers of reflecting elements, e.g., $N_{\rm R}\geq448$. This is because the model mismatch increases with the number of reflecting elements as the near-field propagation becomes dominant. Comparing the findings in Fig. 5 and Fig. 6, we observe that while increasing both $N_{\rm Tx}$  and  $N_{\rm R}$ enhances the SE of the system, their roles in RIS-aided MIMO communication systems are distinct. On one hand, increasing  $N_{\rm Tx}$  at the Tx enhances the spatial DoF, thereby facilitating interference mitigation and enabling higher beamforming gain. On the other hand, as the number of reflecting elements $N_{\rm R}$ increases, the near-field effects become more pronounced, leading to an increase in the model mismatch between the piece-wise near-field channel and the far-field channel models. 

\section{Conclusions}
This paper proposed to adopt a piece-wise near-field channel model for a RIS-aided MIMO system in the presence of CEEs. We considered three channel models (i.e., near-field, piece-wise near-field and far-field) and analyzed the impact of CEEs and model mismatches on the interference distribution. By treating the interference caused by CEEs and model mismatches as noise, we formulated the joint active and passive beamforming design as an optimization problem to maximize the achievable SE taking into account the transmit power constraint for active beamforming matrix and the constant modulus constraint for passive beamforming matrix.  The joint beamforming optimization problem was equivalently transformed into an MSE minimization problem, which was then addressed by the proposed algorithm exploit the BCD and ADPM to handle the constant modulus constraint of RIS elements. We revealed that the adopted piece-wise near-field channel model not only improves the DoF gain but also demonstrates enhanced robustness against CEEs, resulting in higher achievable rates compared to the other channel models. A promising extension of this work is considering  more reasonable parameters (distance and angle) for the error modeling schemes instead of overall channel estimation error modeling, which could lead to tailored channel estimation schemes and robust resource allocation strategies. Another future research direction is to utilize data-driven deep learning networks to select the number of subsurfaces in the piece-wise near-field channel model to achieve the best trade-off between the modeling accuracy and the robustness against CEEs.

\renewcommand{\bibfont}{\footnotesize}
\bibliographystyle{IEEEtran}
\bibliography{myreferences}

\newpage

% your own custom macro containing the $\backslash${\tt{includegraphics}} command to make things
%\bf{If you include a photo:}\vspace{-33pt}
%\begin{IEEEbiography}[{\includegraphics[width=1in,height=1.25in,clip,keepaspectratio]{fig1}}]{Michael Shell}
%Use $\backslash${\tt{begin\{IEEEbiography\}}} and then for the 1st argument use $\backslash${\tt{includegraphics}} to declare and link the author photo.
%Use the author name as the 3rd argument followed by the biography text.
%\end{IEEEbiography}
%\vfill

\end{document}